\documentclass[12pt]{article}
\usepackage{amsmath}
\usepackage{amssymb}
\begin{document}
\newtheorem{thm}{Theorem}[section]
\newtheorem{lem}[thm]{Lemma}
\newtheorem{prop}[thm]{Proposition}
\newtheorem{conj}[thm]{Conjecture}
\newtheorem{cor}[thm]{Corollary}
\newenvironment{dfn}{\medskip\refstepcounter{thm}
\noindent{\bf Definition \thesection.\arabic{thm}\ }}{\medskip}
\newenvironment{cond}{\medskip\refstepcounter{thm}
\noindent{\bf Condition \thesection.\arabic{thm}\ }}{\medskip}
\newenvironment{proof}[1][,]{\medskip\ifcat,#1
\noindent{\it Proof.\ }\else\noindent{\it Proof of #1.\ }\fi}
{\relax\unskip\nobreak ~\hfill$\square$\medskip}
\def\e#1\e{\begin{equation}#1\end{equation}}
\def\ea#1\ea{\begin{align}#1\end{align}}
\def\eq#1{{\rm(\ref{#1})}}
\def\dim{\mathop{\rm dim}}
\def\Re{\mathop{\rm Re}}
\def\Im{\mathop{\rm Im}}
\def\Ker{\mathop{\rm Ker}}
\def\Hol{\mathop{\rm Hol}}
\def\vol{\mathop{\rm vol}}
\def\ind{\mathop{\rm ind}}
\def\Area{\mathop{\rm Area}}
\def\lind{\mathop{\text{\rm l-ind}}}
\def\cal{\mathcal}
\def\bb{\mathbb}
\def\SU{\mathop{\rm SU}}
\def\U{\mathbin{\rm U}}
\def\R{\mathbb{R}}
\def\Z{\mathbb{Z}}
\def\Q{\mathbb{Q}}
\def\C{\mathbb{C}}
\def\CP{\mathbb{CP}}
\def\al{\alpha}
\def\be{\beta}
\def\ka{\kappa}
\def\la{\lambda}
\def\ga{\gamma}
\def\de{\delta}
\def\ep{\epsilon}
\def\th{\theta}
\def\vp{\varphi}
\def\si{\sigma}
\def\ze{\zeta}
\def\De{\Delta}
\def\La{\Lambda}
\def\Om{\Omega}
\def\Ga{\Gamma}
\def\Si{\Sigma}
\def\om{\omega}
\def\d{{\rm d}}
\def\pd{\partial}
\def\db{{\bar\partial}}
\def\ts{\textstyle}
\def\sst{\scriptscriptstyle}
\def\w{\wedge}
\def\lt{\ltimes}
\def\sm{\setminus}
\def\ot{\otimes}
\def\bigot{\bigotimes}
\def\iy{\infty}
\def\ov{\overline}
\def\ra{\rightarrow}
\def\longra{\longrightarrow}
\def\bs{\boldsymbol}
\def\t{\times}
\def\ha{{\textstyle\frac{1}{2}}}
\def\ti{\tilde}
\def\ovB{\,\overline{\!B}}
\def\ovP{\,\overline{\!P}}
\def\ms#1{\vert#1\vert^2}
\def\md#1{\vert #1 \vert}
\def\bmd#1{\bigl\vert #1 \bigr\vert}
\title{On counting special Lagrangian homology 3-spheres}
\author{Dominic Joyce \\ Lincoln College, Oxford, OX1 3DR}
\date{}
\maketitle

\section{Introduction}
\label{cs1}

It is well known \cite{McSa} that one can define {\it Gromov--Witten
invariants} of symplectic manifolds $(M,\om)$ by choosing a generic
compatible metric $g$ and almost complex structure $J$ on $M$, and
counting the $J$-{\it holomorphic curves} $\Si$ in $M$ in a given
homology class, with signs. The invariants are then essentially
independent of the choice of $g,J$. As $J$-holomorphic curves are
{\it calibrated\/ $2$-submanifolds} with respect to the calibration
$\om$ on the Riemannian manifold $(M,g)$, Gromov--Witten invariants
arise by counting the simplest nontrivial kind of calibrated submanifold.

In this paper we shall attempt to define a similar invariant $I$
of generic (almost) Calabi--Yau 3-folds $(M,J,\om,\Om)$ by counting
another kind of calibrated submanifold, {\it special Lagrangian
$3$-submanifolds}, or {\it SL\/ $3$-folds} for short. In fact we
shall consider only special Lagrangian {\it rational homology
$3$-spheres}, as they occur in 0-dimensional moduli spaces, and can be
counted. Let $S(\de)$ be the set of SL homology 3-spheres in $M$ with
homology class $\de$. Suppose $S(\de)$ is finite. Then we shall define
\e
I(\de)=\sum_{N\in S(\de)}w(N),
\label{cs1eq}
\e
where $w(N)$ is a rational `weight function' depending on the 
topology of~$N$.

For $I$ to be interesting it should either be unchanged by smooth
deformations of the underlying almost Calabi--Yau structure
$(J,\om,\Om)$, or else should transform according to some rigid set of
rules as $[\om]$ and $[\Om]$ move about in $H^2(M,\R)$ and $H^3(M,\C)$.
Now whether such an invariant can be made to work, and how it should
be defined, depends very much on the singular behaviour of special
Lagrangian 3-folds, which is not well understood. Therefore much of
this paper is conjectural, though I hope to be able to publish proofs
of many of the conjectures in the next few years, unless someone else
does first.

Here is why singularities of SL 3-folds are important in this problem.
In the moduli space of almost Calabi-Yau structures on $M$ there are 
certain special real hypersurfaces, determined using the homology 
of $M$. At such a hypersurface, some of the SL 3-folds in $M$ will
become singular. An SL 3-fold may exist in $M$ only on one side of
the hypersurface, and become singular at the hypersurface.

More generally, as we approach the hypersurface from one side, one
or more SL 3-folds may collapse down to a singular SL 3-fold, and
then on the other side this singular SL 3-fold is replaced by a
different collection of one or more SL 3-folds. The key question
addressed in this paper is to find a way to define the invariant
$I$ so that it is unchanged, or transforms in a controlled way,
as we cross these hypersurfaces and the set of SL 3-folds that
we are `counting' changes.

In \S\ref{cs3}--\S\ref{cs4} and \S\ref{cs5}--\S\ref{cs6} we study
two kinds of singularity that SL 3-folds can develop, modelled
respectively on a $T^2$-cone in $\C^3$, and the union of two SL
3-planes $\R^3$ in $\C^3$. We use these transitions to calculate
identities which the weight function $w(N)$ in \eq{cs1eq} must
satisfy for $I$ to be invariant, or transform nicely, as we pass
through the hypersurface. It turns out that the simple weight
function $w(N)=\bmd{H_1(N,\Z)}$ satisfies these identities.

Motivated by this, we formulate a conjecture, Conjecture
\ref{cs7conj}, giving a partial definition of $I$, and a partial
statement of the transformation law it should satisfy under
deformation. A full definition and transformation law (if the
invariant works at all) will have to await a better understanding
of singular SL 3-folds. We conclude in \S\ref{cs76} with a discussion
of the relationship of the invariants to String Theory. The author
believes that they count objects of significance in String Theory,
namely {\it isolated\/ $3$-branes}, and that they should play a
r\^ole in Mirror Symmetry of Calabi--Yau 3-folds.
\medskip

\noindent{\it Acknowledgements.} I would like to thank Bobby
Acharya, Edward Witten, Simon Donaldson, and Mark Gross for
helpful conversations.

\section{Special Lagrangian geometry}
\label{cs2}

We now introduce the idea of special Lagrangian submanifolds
(SL $m$-folds), in two different geometric contexts. First,
in \S\ref{cs21}, we define SL $m$-folds in $\C^m$. Then
\S\ref{cs22} discusses SL $m$-folds in {\it almost Calabi--Yau
$m$-folds}, compact K\"ahler manifolds equipped with a
holomorphic volume form which generalize the idea of
Calabi--Yau manifolds. Finally, section \ref{cs23} considers
the {\it singularities} of SL $m$-folds. The principal references
for this section are Harvey and Lawson \cite{HaLa} and the
author~\cite{Joyc7}.

\subsection{Special Lagrangian submanifolds in $\C^m$}
\label{cs21}

We begin by defining {\it calibrations} and {\it calibrated 
submanifolds}, following Harvey and Lawson~\cite{HaLa}.

\begin{dfn} Let $(M,g)$ be a Riemannian manifold. An {\it oriented
tangent $k$-plane} $V$ on $M$ is a vector subspace $V$ of
some tangent space $T_xM$ to $M$ with $\dim V=k$, equipped
with an orientation. If $V$ is an oriented tangent $k$-plane
on $M$ then $g\vert_V$ is a Euclidean metric on $V$, so 
combining $g\vert_V$ with the orientation on $V$ gives a 
natural {\it volume form} $\vol_V$ on $V$, which is a 
$k$-form on~$V$.

Now let $\vp$ be a closed $k$-form on $M$. We say that
$\vp$ is a {\it calibration} on $M$ if for every oriented
$k$-plane $V$ on $M$ we have $\vp\vert_V\le \vol_V$. Here
$\vp\vert_V=\al\cdot\vol_V$ for some $\al\in\R$, and 
$\vp\vert_V\le\vol_V$ if $\al\le 1$. Let $N$ be an 
oriented submanifold of $M$ with dimension $k$. Then 
each tangent space $T_xN$ for $x\in N$ is an oriented
tangent $k$-plane. We say that $N$ is a {\it calibrated 
submanifold} if $\vp\vert_{T_xN}=\vol_{T_xN}$ for all~$x\in N$.
\label{cs2def1}
\end{dfn}

It is easy to show that calibrated submanifolds are automatically
{\it minimal submanifolds} \cite[Th.~II.4.2]{HaLa}. Here is the 
definition of special Lagrangian submanifolds in $\C^m$, taken
from~\cite[\S III]{HaLa}.

\begin{dfn} Let $\C^m$ have complex coordinates $(z_1,\dots,z_m)$, 
and define a metric $g$, a real 2-form $\om$ and a complex $m$-form 
$\Om$ on $\C^m$ by
\e
\begin{split}
g=\ms{\d z_1}+\cdots+\ms{\d z_m},\quad
\om&=\ts\frac{i}{2}(\d z_1\w\d\bar z_1+\cdots+\d z_m\w\d\bar z_m),\\
\text{and}\quad\Om&=\d z_1\w\cdots\w\d z_m.
\end{split}
\label{cs2eq1}
\e
Then $\Re\Om$ and $\Im\Om$ are real $m$-forms on $\C^m$. Let $L$
be an oriented real submanifold of $\C^m$ of real dimension $m$. We
say that $L$ is a {\it special Lagrangian submanifold\/} of $\C^m$,
or {\it SL\/ $m$-fold}\/ for short, if $L$ is calibrated with respect
to $\Re\Om$, in the sense of Definition~\ref{cs2def1}.
\label{cs2def2}
\end{dfn}

Harvey and Lawson \cite[Cor.~III.1.11]{HaLa} give the following
alternative characterization of special Lagrangian submanifolds:

\begin{prop} Let\/ $L$ be a real $m$-dimensional submanifold 
of $\C^m$. Then $L$ admits an orientation making it into an
SL submanifold of\/ $\C^m$ if and only if\/ $\om\vert_L\equiv 0$ 
and\/~$\Im\Om\vert_L\equiv 0$.
\label{cs2prop1}
\end{prop}

An $m$-dimensional submanifold $L$ in $\C^m$ is called {\it Lagrangian} 
if $\om\vert_L\equiv 0$. Thus special Lagrangian submanifolds are 
Lagrangian submanifolds satisfying the extra condition that 
$\Im\Om\vert_L\equiv 0$, which is how they get their name.

\subsection{Almost Calabi--Yau $m$-folds and SL $m$-folds} 
\label{cs22}

We shall define special Lagrangian submanifolds not just in
Calabi--Yau manifolds, as usual, but in the much larger
class of {\it almost Calabi--Yau manifolds}.

\begin{dfn} Let $m\ge 2$. An {\it almost Calabi--Yau $m$-fold}, or
{\it ACY\/ $m$-fold}\/ for short, is a quadruple $(M,J,\om,\Om)$ 
such that $(M,J)$ is a compact $m$-dimensional complex manifold,
$\om$ is the K\"ahler form of a K\"ahler metric $g$ on $M$, and
$\Om$ is a non-vanishing holomorphic $(m,0)$-form on~$M$.

We call $(M,J,\om,\Om)$ a {\it Calabi--Yau $m$-fold}, or {\it CY\/ 
$m$-fold}\/ for short, if in addition $\om$ and $\Om$ satisfy
\e
\om^m/m!=(-1)^{m(m-1)/2}(i/2)^m\Om\w\bar\Om.
\label{cs2eq2}
\e
Then for each $x\in M$ there exists an isomorphism $T_xM\cong\C^m$
that identifies $g_x,\om_x$ and $\Om_x$ with the flat versions
$g,\om,\Om$ on $\C^m$ in \eq{cs2eq1}. Furthermore, $g$ is Ricci-flat
and its holonomy group is a subgroup of~$\SU(m)$.
\label{cs2def3}
\end{dfn}

This is not the usual definition of a Calabi--Yau manifold, but
is essentially equivalent to it.

\begin{dfn} Let $(M,J,\om,\Om)$ be an almost Calabi--Yau $m$-fold, and
$N$ an oriented real $m$-dimensional submanifold of $M$. Fix $\th\in\R$.
We call $N$ a {\it special Lagrangian submanifold}, or {\it SL $m$-fold}
for short, with {\it phase} ${\rm e}^{i\th}$ if
\e
\om\vert_N\equiv 0 \quad\text{and}\quad 
(\sin\th\,\Re\Om-\cos\th\,\Im\Om)\vert_N\equiv 0,
\label{cs2eq3}
\e
and $\cos\th\,\Re\Om+\sin\th\,\Im\Om$ is a positive $m$-form on 
the oriented $m$-fold~$N$.
\label{cs2def4}
\end{dfn}

Again, this is not the usual definition of special Lagrangian
submanifold, but is essentially equivalent to it. Compared to
Definition \ref{cs2def1} we have introduced two changes:
\begin{itemize}
\item[(a)] the definition now involves a {\it phase}
${\rm e}^{i\th}$, and
\item[(b)] SL $m$-folds are defined by the {\it vanishing of forms}\/
$\om$ and $\sin\th\,\Re\Om-\cos\th\,\Im\Om$, rather than as calibrated
submanifolds.
\end{itemize}
The following easy lemma relates the phase of a compact SL $m$-fold
to its homology class.

\begin{lem} Let\/ $(M,J,\om,\Om)$ be an almost Calabi--Yau $m$-fold,
and\/ $N$ a compact SL $m$-fold in $M$ with phase ${\rm e}^{i\th}$.
Then
\e
[\Om]\cdot[N]=R{\rm e}^{i\th}, \quad\text{where}\quad
R=\int_N\cos\th\,\Re\Om+\sin\th\,\Im\Om>0,
\label{cs2eq4}
\e
$[\Om]\in H^m(M,\C)$ and\/ $[N]\in H_m(M,\Z)$. Thus the homology
class $[N]$ determines the phase ${\rm e}^{i\th}$ of\/~$N$.
\label{cs2lem}
\end{lem}

If we study only SL $m$-folds $N$ in a fixed homology class in $M$,
then by replacing $\Om$ by ${\rm e}^{-i\th}\Om$ we can suppose that
$N$ has phase 1, and so avoid introducing phases. However, we will
later need to consider several SL 3-folds $N_1,N_2,N_3$ in different
homology classes in a Calabi--Yau 3-fold $M$, with different phases,
so we can discuss what happens to SL 3-folds as we vary $[\Om]\in
H^3(M,\C)$. This is why we introduced change~(a).

Using the analogue of Proposition \ref{cs2prop1} one can show that
if $(M,J,\om,\Om)$ is a Calabi--Yau manifold, then $N$ is special
Lagrangian with phase ${\rm e}^{i\th}$ if and only if it is calibrated
w.r.t.\ $\cos\th\,\Re\Om+\sin\th\,\Im\Om$. More generally
\cite[\S 9.4]{Joyc7}, SL $m$-folds in an almost Calabi--Yau
$m$-fold are calibrated w.r.t.\ $\cos\th\,\Re\Om+\sin\th\,\Im\Om$,
but for a suitably {\it conformally rescaled} metric~$g$.

Thus, we can give an equivalent definition of SL $m$-folds in terms
of calibrated geometry, and change (b) above is only cosmetic.
Nonetheless, in the author's view the definition of SL $m$-folds in
terms of the vanishing of closed forms is more fundamental than the
definition in terms of calibrated geometry, and so should be taken
as the primary definition.

The {\it deformation theory} of special Lagrangian submanifolds
was studied by McLean \cite[\S 3]{McLe}, who proved the following
result in the Calabi--Yau case with phase 1. The extension to the
ACY case is described in~\cite[\S 9.5]{Joyc7}.

\begin{thm} Let\/ $(M,J,\om,\Om)$ be an almost Calabi--Yau $m$-fold,
and\/ $N$ a compact SL\/ $m$-fold in $M$. Then the moduli space
${\cal M}_N$ of special Lagrangian deformations of\/ $N$ is a smooth
manifold of dimension $b^1(N)$, the first Betti number of\/ $N$. All
elements of\/ ${\cal M}_N$ have the same phase ${\rm e}^{i\th}$,
given by $[\Om]\cdot[N]=R{\rm e}^{i\th}$ for~$R>0$.
\label{cs2thm1}
\end{thm}

Using similar methods one can prove~\cite[\S 9.3, \S 9.5]{Joyc7}:

\begin{thm} Let\/ $\bigl\{(M,J_t,\om_t,\Om_t):t\in(-\ep,\ep)\bigr\}$ be
a smooth\/ $1$-parameter family of almost Calabi--Yau $m$-folds. Let\/
$N_0$ be a compact SL\/ $m$-fold in $(M,J_0,\om_0,\Om_0)$, and suppose
$[\om_t\vert_{N_0}]=0$ in $H^2(N_0,\R)$ for all\/ $t\in(-\ep,\ep)$.
Then $N_0$ extends to a smooth\/ $1$-parameter family $\bigl\{N_t:
t\in(-\de,\de)\bigr\}$, where $0<\de\le\ep$ and\/ $N_t$ is a compact
SL\/ $m$-fold in $(M,J_t,\om_t,\Om_t)$, with phase ${\rm e}^{i\th_t}$
determined by $[\Om_t]\cdot[N_0]=R_t{\rm e}^{i\th_t}$ for~$R_t>0$.
\label{cs2thm2}
\end{thm}

Now suppose that $(M,J,\om,\Om)$ is an almost Calabi--Yau 3-fold,
and $N$ is a {\it special Lagrangian (rational) homology $3$-sphere}
in $M$. Then $H^1(N,\R)=H^2(N,\R)=0$. Thus by Theorem \ref{cs2thm1}
the moduli space ${\cal M}_N$ has dimension 0, so that $N$ is
{\it rigid} (that is, it admits no nontrivial deformations as an SL
3-fold in $M$). Also, as $H^2(N,\R)=0$ the condition $[\om_t\vert_N]=0$
in Theorem \ref{cs2thm2} holds automatically for any family of
deformations $(J_t,\om_t,\Om_t)$ of the almost Calabi--Yau
structure $(J,\om,\Om)$ on $M$. Thus, Theorem \ref{cs2thm2} shows
that $N$ is {\it stable} under small deformations of $(J,\om,\Om)$,
giving:

\begin{cor} Let\/ $(M,J,\om,\Om)$ be an almost Calabi--Yau $3$-fold,
and\/ $N$ a special Lagrangian homology $3$-sphere in $M$. Then $N$
is rigid, and stable under small deformations of the almost
Calabi--Yau structure $(J,\om,\Om)$ on~$M$.
\label{cs2cor}
\end{cor}

One moral of this is that special Lagrangian homology 3-spheres
in an almost Calabi--Yau 3-fold may be a good thing to count, as
they are isolated and persistent under small deformations. We will
discuss this in~\S\ref{cs7}.

\subsection{Singularities of SL $m$-folds}
\label{cs23}

The author's series of papers
\cite{Joyc1,Joyc2,Joyc3,Joyc4,Joyc5,Joyc6,Joyc7} is mainly concerned
with the study of singularities of SL $m$-folds in $\C^m$ and in
almost Calabi--Yau manifolds. This is done both by the construction
of many examples of singular SL $m$-folds in $\C^m$, and also through
the development of a (partly conjectural) picture of how families of
nonsingular SL $m$-folds can become singular, particularly when the
underlying almost Calabi--Yau manifold is assumed to be {\it generic}.

We shall now briefly summarize some definitions and conjectures
taken from \cite[\S 10]{Joyc7}. The author has sketch proofs for
the conjectures (at least in complex dimension $m<6$) and hopes
to publish full proofs fairly soon. For more details and motivation,
see \cite{Joyc7}. Here are some definitions to do with {\it special
Lagrangian cones} in~$\C^m$.

\begin{dfn} A (singular) SL $m$-fold $C$ in $\C^m$ is called a
{\it cone} if $C=tC$ for all $t>0$, where $tC=\{tx:x\in C\}$.
Let $C$ be an SL cone in $\C^m$. Then either $C$ is an $m$-plane
$\R^m$ in $\C^m$, or $C$ is singular at 0. We are interested
primarily in SL cones $C$ in which 0 is the only singular point,
that is, in which 0 is an {\it isolated singularity}. Then
$\Si=C\cap{\cal S}^{2m-1}$ is a compact, nonsingular
$(m-1)$-submanifold of ${\cal S}^{2m-1}$. We define the
{\it number of ends at infinity} of $C$ to be the number $k$
of connected components of~$\Si$.

Let $C$ be an SL cone in $\C^m$ with an isolated singularity at 0,
and let $\Si=C\cap{\cal S}^{2m-1}$. Regard $\Si$ as a compact
Riemannian manifold, with metric induced from the round metric in
${\cal S}^{2m-1}$. Let $\De=\d^*\d$ be the Laplacian on functions
on $\Si$. Define the {\it Legendrian index} $\lind(C)$ to be the
number of eigenvalues of $\De$ in $(0,2m)$, counted with multiplicity.

Let the connected components of $\Si$ be $\Si_1,\ldots,\Si_k$. Define
the cone $C$ to be {\it rigid} if for each $j=1,\ldots,k$, the
eigenspace of $\De$ on $\Si_j$ with eigenvalue $2m$ has dimension
$\dim SU(m)-\dim G_j$, where $G_j$ is the Lie subgroup of $\SU(m)$
preserving~$\Si_j$.
\label{cs2def5}
\end{dfn}

The point of these definitions is that the Legendrian index of
$C$ is that $\Si$ is a {\it minimal Legendrian submanifold} in
${\cal S}^{2m-1}$, and is thus a stationary point of the area
functional amongst all Legendrian submanifolds in ${\cal S}^{2m-1}$.
The {\it Legendrian index} is the index of this stationary point.
The cone $C$ is the union of one-ended cones $C_1,\ldots,C_k$
intersecting at 0, and $C$ is {\it rigid} if all infinitesimal
deformations of $C_j$ as an SL cone come from infinitesimal
rotations of $C_j$ by $\SU(m)$ matrices, for each~$j$.

Now SL cones are important because they are local models for
the simplest kind of singularities of SL $m$-folds in almost
Calabi--Yau $m$-folds. To understand how singular SL $m$-folds
modelled upon an SL cone $C$ in $\C^m$ can arise as limits of
nonsingular SL $m$-folds, we need to consider SL $m$-folds $L$
in $\C^m$ asymptotic to $C$ at infinity.

\begin{dfn} Let $C$ be an SL cone in $\C^m$ with isolated
singularity at 0, and let $\Si=C\cap{\cal S}^{2m-1}$, so that
$\Si$ is a compact, nonsingular $(m-1)$-manifold. Let $h$ be
the metric on $\Si$ induced by the metric $g$ on $\C^m$, and $r$
the radius function on $\C^m$. Define $\iota:\Si\t(0,\iy)\ra\C^m$
by $\iota(\si,r)=r\si$. Then the image of $\iota$ is $C\sm\{0\}$,
and $\iota^*(g)=r^2h+\d r^2$ is the cone metric on~$C\sm\{0\}$.

Let $L$ be a closed, nonsingular SL $m$-fold in $\C^m$. We call $L$
{\it Asymptotically Conical (AC)} with cone $C$ if there exists a
compact subset $K\subset L$ and a diffeomorphism $\phi:\Si\t(R,\iy)
\ra L\sm K$ for some $R>0$, such that $\md{\phi-\iota}=o(r)$ and
$\bmd{\nabla^k(\phi-\iota)}=o(r^{1-k})$ as $r\ra\iy$ for $k=1,2,\ldots$,
where $\nabla$ is the Levi-Civita connection of the cone metric
$\iota^*(g)$, and $\md{\,.\,}$ is computed using~$\iota^*(g)$.
\label{cs2def6}
\end{dfn}

In \cite[\S 10]{Joyc7} this notion of Asymptotically Conical is
referred to as {\it weakly Asymptotically Conical}, to distinguish
it from a second class of {\it strongly Asymptotically Conical\/}
SL $m$-folds which converge to $C$ to order $O(r^{-1})$ rather than
$o(r)$. However, we will not need the idea of strongly AC SL $m$-folds
in this paper. The following conjecture \cite[Conj.~10.3]{Joyc7} is
the analogue of Theorem \ref{cs2thm1} for AC SL $m$-folds.

\begin{conj} Let\/ $L$ be an AC SL\/ $m$-fold in $\C^m$,
with cone $C$, and let\/ $k$ be the number of ends of\/ $C$
at infinity. Then the moduli space ${\cal M}_{\sst L}$ of AC
SL\/ $m$-folds in $\C^m$ with cone $C$ is near $L$ a smooth
manifold of dimension~$b^1(L)+k-1+\lind(C)$.
\label{cs2conj1}
\end{conj}

Our next conjecture \cite[Conj.~10.7]{Joyc7} is a first approximation
to the kinds of deformation results the author expects to hold for
singular SL $m$-folds in almost Calabi--Yau $m$-folds.

\begin{conj} Let\/ $C$ be a rigid SL cone in $\C^m$ with an isolated
singularity at\/ $0$ and\/ $k$ ends at infinity, and\/ $L$ be an AC
SL\/ $m$-fold in $\C^m$ with cone $C$. Let\/ $(M,J,\om,\Om)$ be a
generic almost Calabi--Yau $m$-fold, and\/ $\cal M$ a connected moduli
space of compact nonsingular SL\/ $m$-folds $N$ in~$M$.

Suppose that at the boundary of\/ $\cal M$ there is a moduli space 
${\cal M}_C$ of compact, singular SL\/ $m$-folds with one isolated 
singular point modelled on the cone $C$, which arise as limits of
SL\/ $m$-folds in $\cal M$ by collapsing AC SL\/ $m$-folds with the
topology of\/ $L$. Then
\e
\dim{\cal M}=\dim{\cal M}_C+b^1(L)+k-1+\lind(C)-2m.
\label{cs2eq5}
\e
\label{cs2conj2}
\end{conj}

Suppose we have a suitably generic almost Calabi--Yau $m$-fold $M$
and a compact, singular SL $m$-fold $N_0$ in $M$, which is the limit
of a family of compact nonsingular SL $m$-folds $N$ in $M$. We
(loosely) define the {\it index} of the singularities of $N_0$ to
be the codimension of the family of singular SL $m$-folds with
singularities like those of $N_0$ in the family of nonsingular SL
$m$-folds $N$. Thus, in the situation of Conjecture \ref{cs2conj2},
the index of the singularities is~$b^1(L)+k-1+\lind(C)-2m$.

More generally, one can work not just with a fixed generic almost
Calabi--Yau $m$-fold, but with a {\it generic family} of almost
Calabi--Yau $m$-folds. So, for instance, if we have a generic
$k$-dimensional family of almost Calabi--Yau $m$-folds $M$, and in 
each $M$ we have an $l$-dimensional family of SL $m$-folds, then 
in the total $(k\!+\!l)$-dimensional family of SL $m$-folds we are 
guaranteed to meet singularities of index at most~$k\!+\!l$.

Now later in the paper we shall study the behaviour of 0-dimensional
families of SL 3-folds in generic 1-dimensional families of almost
Calabi--Yau 3-folds. In such families we will only meet singularities
of index 1. This is a very useful fact, as it means there will be
only a few kinds of singular behaviour to worry about in determining
how the invariants we define behave under deformations of the
underlying almost Calabi--Yau 3-fold.

One important reason we have chosen to work in almost Calabi--Yau
manifolds, rather than just in Calabi--Yau manifolds, is that almost
Calabi--Yau manifolds occur in {\it infinite-dimensional\/} families.
Thus, taking the underlying almost Calabi--Yau manifold to be generic
is a very powerful assumption, and should simplify the singular
behaviour of SL $m$-folds considerably. (For instance, one can
argue that a compact SL 3-fold in a generic almost Calabi--Yau
3-fold has at most finitely many singular points.) However,
Calabi--Yau manifolds only occur in finite-dimensional
families, and so working in a generic Calabi--Yau manifold is not
that strong an assumption, and probably will not help very much.

\section{A model degeneration of SL 3-folds}
\label{cs3}

We now define an explicit SL cone $L_0$ in $\C^3$ and three families
of AC SL 3-folds $L^a_t$ in $\C^3$ with cone $L_0$, and analyze them
in the framework of~\S\ref{cs23}.

\subsection{Three families of SL 3-folds in $\C^3$}
\label{cs31}

Let $G$ be the group $\U(1)^2$, acting on $\C^3$ by
\e
({\rm e}^{i\th_1},{\rm e}^{i\th_2}):(z_1,z_2,z_3)\mapsto
({\rm e}^{i\th_1}z_1,{\rm e}^{i\th_2}z_2,{\rm e}^{-i\th_1-i\th_2}z_3)
\quad\text{for $\th_1,\th_2\in\R$.}
\label{cs3eq1}
\e
All the $G$-invariant special Lagrangian 3-folds in $\C^3$ were
written down explicitly by Harvey and Lawson \cite[\S III.3.A]{HaLa},
and studied in more detail in \cite[Ex.~5.1]{Joyc1} and
\cite[\S 4]{Joyc4}. Here are some examples of $G$-invariant SL
3-folds which will be important in what follows.

\begin{dfn} Define a subset $L_0$ in $\C^3$ by
\e
\begin{split}
L_0=\bigl\{(z_1,z_2&,z_3)\in\C^3:\ms{z_1}=\ms{z_2}=\ms{z_3},\\
&\Im(z_1z_2z_3)=0,\quad \Re(z_1z_2z_3)\ge 0\bigr\}.
\end{split}
\label{cs3eq2}
\e
Then $L_0$ is a {\it special Lagrangian cone} on $T^2$, invariant
under the Lie subgroup $G$ of $\SU(3)$ given in \eq{cs3eq1}. Let
$t>0$, and define
\ea
\begin{split}
L^1_t=\bigl\{(z_1,z_2,z_3)\in\C^3:\,&
\ms{z_1}-t=\ms{z_2}=\ms{z_3},\\
&\Im(z_1z_2z_3)=0,\quad \Re(z_1z_2z_3)\ge 0\bigr\}.
\end{split}
\label{cs3eq3}\\
\begin{split}
L^2_t=\bigl\{(z_1,z_2,z_3)\in\C^3:\,&
\ms{z_1}=\ms{z_2}-t=\ms{z_3},\\
&\Im(z_1z_2z_3)=0,\; \Re(z_1z_2z_3)\ge 0\bigr\},
\end{split}
\label{cs3eq4}\\
\begin{split}
L^3_t=\bigl\{(z_1,z_2,z_3)\in\C^3:\,&
\ms{z_1}=\ms{z_2}=\ms{z_3}-t,\\
&\Im(z_1z_2z_3)=0,\; \Re(z_1z_2z_3)\ge 0\bigr\}.
\end{split}
\label{cs3eq5}
\ea
Then it can be shown that each $L^a_t$ is a $G$-invariant, nonsingular,
embedded special Lagrangian 3-submanifold in $\C^3$ diffeomorphic to
${\cal S}^1\t\R^2$, which is {\it Asymptotically Conical} in the sense
of Definition \ref{cs2def5}, with cone~$L_0$.
\label{cs3def}
\end{dfn}

Thus the $L^a_t$ for $a=1,2,3$ are {\it three different} families
of AC SL 3-folds in $\C^3$ asymptotic to the same SL cone $L_0$,
each family depending on a real parameter $t>0$. Define subsets
$D^1_t$ and $\ga^1_t$ in $\C^3$ for $a>0$ by
\e
\ga^1_t=\bigl\{(t^{1/2}\,{\rm e}^{i\th},0,0):
\th\in\R\bigr\},\;\>
D^1_t=\bigl\{(z_1,0,0):z_1\in\C,\;\> \ms{z_1}\le t\bigr\}.
\label{cs3eq6}
\e
Then $\ga^1_t$ is a smooth, oriented ${\cal S}^1$ in $L^1_t$, and
$D^1_t$ is a closed, oriented holomorphic disc in $\C^3$ with
area $\pi t$ and boundary $\ga^1_t$. The homology class of 
$\ga^1_t$ generates $H_1(L^1_t,\Z)\cong\Z$. There are similar 
holomorphic discs $D^2_t$ with boundary $\ga^2_t$ in $L^2_t$, 
and $D^3_t$ with boundary $\ga^3_t$ in~$L^3_t$.

\subsection{Using $L_0$ and the $L^a_t$ as local models}
\label{cs32}

Next we apply the ideas of \S\ref{cs23} to $L_0$ and the~$L^a_t$.

\begin{lem} In the notation of\/ \S\ref{cs23}, the SL cone $L_0$
and AC SL\/ $3$-folds $L^a_t$ satisfy $k=1$, $b^1(L^a_t)=1$
and\/ $\lind(L_0)=6$, and\/ $L_0$ is rigid.
\label{cs3lem}
\end{lem}

\begin{proof} As $L_0$ is a $T^2$-cone we have $k=1$, and
as $L^a_t$ is diffeomorphic to ${\cal S}^1\t\R^2$ we have
$b^1(L^a_t)=1$. It is not difficult to show that the metric
on $\Si\cong T^2$ is isometric to the quotient of $\R^2$ with
its flat Euclidean metric by the lattice $\Z^2$ with basis
$2\pi\bigl(\frac{\sqrt{2}}{\sqrt{3}},0\bigl)$,
$2\pi\bigl(\frac{1}{\sqrt{6}},\frac{1}{\sqrt{2}}\bigr)$.
The eigenvectors of $\De$ on $\Si$ lift to functions of the
form $\sin(\al x+\be y),\cos(\al x+\be y)$ on $\R^2$ which
are invariant under lattice translations.

Calculation shows that the only eigenvalue of $\De$ in $(0,6)$
is 2, with multiplicity 6, and so $\lind(L_0)=6$. Similarly, $\De$
has eigenvalue 6 with multiplicity 6, and as $\dim\SU(3)=8$ and 
the subgroup $G$ of $\SU(3)$ preserving $L_0$ is $\U(1)^2$ with
dimension 2, we see that $L_0$ is rigid as~$6=8-2$.
\end{proof}

Assuming Conjecture \ref{cs2conj1}, we see that the moduli
space of AC SL 3-folds $L$ with cone $L_0$ and $b^1(L)=1$ is
1-dimensional. It follows that any such $L$ must be $G$-invariant,
as otherwise the $G$-orbit of $L$ and its rescalings would fill
out a moduli space of larger dimension. Thus any such $L$ is
one of the $L^a_t$, and we have:

\begin{cor} Suppose Conjecture \ref{cs2conj1} holds when $m=3$.
Then the moduli space of AC SL\/ $3$-folds $L$ in $\C^3$ with
cone $L_0$ and\/ $b^1(L)=1$ is~$\{L^a_t:a=1,2,3$,~$t>0\}$.
\label{cs3cor1}
\end{cor}

Assuming Conjecture \ref{cs2conj2}, we deduce:

\begin{cor} Suppose Conjecture \ref{cs2conj2} holds when $m=3$.
Let\/ $(M,J,\om,\Om)$ be a generic almost Calabi--Yau $3$-fold,
and\/ $\cal M$ a connected moduli space of compact nonsingular
SL\/ $3$-folds $N$ in $M$. Suppose that at the boundary of\/
$\cal M$ there is a moduli space ${\cal M}_0$ of compact, singular
SL\/ $3$-folds with one isolated singular point modelled on the
cone $L_0$, which arise as limits of SL\/ $3$-folds in $\cal M$
by collapsing AC SL\/ $3$-folds of the form $L^a_t$.
Then~$\dim{\cal M}=\dim{\cal M}_0+1$.
\label{cs3cor2}
\end{cor}

Thus, in the sense discussed in \S\ref{cs23}, singularities of SL
3-folds modelled on the SL cone $L_0$ should have {\it index one}.
That is, they should occur in codimension 1 in families of SL
3-folds in a generic almost Calabi--Yau 3-fold, or more generally
in generic families of almost Calabi--Yau 3-folds.

\subsection{Interpretation in terms of holomorphic discs}
\label{cs33}

Here is an informal way to understand why singularities of SL 3-folds
modelled on $L_0$ should have index one, in terms of {\it holomorphic
discs}. Given a symplectic manifold $(M,\om)$ with compatible almost
complex structure $J$, and a Lagrangian submanifold $N$ in $M$,
one can consider {\it $J$-holomorphic discs} $D$ in $M$ with
boundary $\pd D$ in~$N$.

The behaviour of such $J$-holomorphic discs is well understood,
because of their application in the Floer homology of Lagrangian
submanifolds. It is known (see for instance \cite[Th.~3.11]{FOOO})
that if $J$ and $N$ are sufficiently generic, then moduli spaces of
$J$-holomorphic discs are smooth manifolds, with dimension given
by a topological formula.

In our case, as $N$ is special Lagrangian the `Maslov index' of
$D$ is automatically 0, and as $m=3$ the formula gives dimension
0 for the moduli space of holomorphic discs. The conclusion is
that if $(M,J,\om,\Om)$ is a generic almost Calabi--Yau 3-fold
and $N$ an SL 3-fold in $M$, then holomorphic discs $D$ with
boundary $\pd D$ in $N$ are expected to be both {\it isolated}
and {\it stable}, so that they persist under small deformations
of $(J,\om,\Om)$ and~$N$.

In the situation of Corollary \ref{cs3cor2}, suppose that $N_0$
is a singular SL 3-fold in ${\cal M}_0$ with singular point $x$,
and $N$ is an SL 3-fold in $\cal M$ close to $N_0$. Then we expect
that $N$ should be modelled near $x$ on some $L^a_t$ for small $t>0$.
(This will be made more precise in \S\ref{cs4}.) Since there is
a holomorphic disc $D^a_t$ in $\C^3$ with boundary in $L^a_t$, and
such discs are isolated and stable under small perturbations, we
expect that there should exist a holomorphic disc $D$ in $M$ with
boundary in $N$ modelled on~$D^a_t$.

As $D$ is calibrated with respect to $\om$, its area is $\int_D\om$.
Thus, the area of $D$ is given {\it topologically} in terms of relative
homology and cohomology by the formula $\Area(D)=[\om]_{M;N}\cdot
[D]_{M;N}$, where $[\om]_{M;N}\in H^2(M;N,\R)$ and~$[D]_{M;N}\in
H_2(M;N,\Z)$. But also, the area of $D$ is necessarily {\it positive}.
When the area shrinks to zero, the boundary ${\cal S}^1$ of $D$ in $N$
is collapsed to a point, and $N$ develops a singularity, a $T^2$-cone.

Thus, we have the following picture. Regard $[D]_{M,N}$ as a fixed
class $\be\in H_2(M;N,\Z)$. As we continuously vary $(M,J,\om,\Om)$
and $N$, the class $[\om]_{M;N}$ varies in $H^2(M;N,\R)$. On
$\cal M$ we have $[\om]_{M;N}\cdot\be>0$, and ${\cal M}_0$ is the
boundary where $[\om]_{M;N}\cdot\be=0$. So, singularities modelled
on the cone $L_0$ should occur at the zeroes of one topologically
determined real function $[\om]_{M;N}\cdot\be$ on the moduli space
of SL 3-folds, which explains why they happen in codimension one.

The author expects that the SL cone $L_0$ in $\C^3$ is the generic
local model for singularities of SL 3-folds which occur when the
area of a holomorphic disc $D$ with boundary in an SL 3-fold $N$
shrinks to zero.

\section{Topological behaviour of these singularities}
\label{cs4}

In \S\ref{cs3} we argued that singularities modelled on the SL cone
$L_0$ are of {\it index one}, in the sense of \S\ref{cs23}. Therefore
we expect singularities modelled on $L_0$ to occur in codimension one
in families of SL 3-folds in generic families of almost Calabi--Yau
3-folds. We shall consider such a family, and investigate the
topological consequences.

\subsection{SL 3-folds with $T^2$-cone singularities}
\label{cs41}

In the following condition we set out the situation we want to study,
an SL 3-fold in an almost Calabi--Yau 3-fold with one singular point
modelled on $L_0$, and establish some notation.

\begin{cond} Let $(M,J_0,\om_0,\Om_0)$ be an almost Calabi--Yau 3-fold,
and $N_0$ a compact, embedded, singular SL 3-fold in $M$ with phase
${\rm e}^{i\th}$ and one singular point at $x\in M$, locally modelled
on $L_0$. To be more precise, let $\rho>0$ be small, $B_\rho$ and
$\ovB_\rho$ be the open and closed balls of radius $\rho$ about 0 in
$\C^3$, and $(J',\om',\Om')$ the Euclidean Calabi--Yau structure on
$\ovB_\rho\subset\C^3$. Suppose there exists a $C^1$ embedding
$\Phi_0:\ovB_\rho\ra M$ which is smooth except perhaps at 0, such that
\begin{itemize}
\item[(a)] $\Phi_0(0)=x$ and $\Phi_0^*(N_0)=L_0\cap\ovB_\rho$;
\item[(b)] $\Phi_0^*(\om_0)=\om'$; and
\item[(c)] $\Phi_0^*(J_0)\approx J'$ and $\Phi_0^*(\Om_0)\approx
F{\rm e}^{i\th}\Om'$ near $0\in \ovB_\rho$, for some $F>0$.
\end{itemize}
Suppose also that~$H_1(N_0,\R)=\{0\}$.
\label{cs4cond}
\end{cond}

Here are some remarks on this condition:
\begin{itemize}
\item As our conclusions will be conjectural, we will not worry very
much about the details here -- for instance, exactly what sort of
approximation is required in part (c), or how many derivatives of
$\Phi_0$ exist at~$x$.
\item We have chosen the coordinate system $\Phi_0$ to equate the
symplectic structures $\om_0$ on $M$ and $\om'$ on $\C^3$. This is
possible by the Darboux Lemma. As there are so many symplectomorphisms
we can also arrange that $(\Phi_0)^*(N_0)=L_0\cap\ovB_\rho$, possibly
at the cost of $\Phi_0$ not being smooth at 0. However, $\Phi_0$ will
in general not be holomorphic, so we can only assume that $\Phi_0^*
(J_0)\approx J'$ and $\Phi_0^*(\Om_0)\approx F{\rm e}^{i\th}\Om'$.

Here the factor ${\rm e}^{i\th}$ is to compensate for the fact that
$N_0$ has phase ${\rm e}^{i\th}$ and $L_0$ has phase 1, and $F$ is
to allow for the fact that as $(M,J_0,\om_0,\Om_0)$ is only an
{\it almost}\/ Calabi--Yau 3-fold, so $(J_0,\om_0,\Om_0)$ is
not necessarily isomorphic to $(J',\om',\Om')$ at $x$, but only
to $(J',\om',F\Om')$ for some~$F>0$.

One could instead choose the $\Phi_0$ to be holomorphic coordinates,
with $\Phi_0^*(J_0)=J'$ and $\Phi_0^*(\Om_0)=\Om'$. But then the best
we could hope for would be $\Phi_0^*(\om_0)\approx F\om'$
and~$\Phi_0^*(N_0)\approx L_0\cap\ovB_\rho$.

\item The assumptions that $N_0$ is embedded and $H_1(N_0,\R)=\{0\}$
are to simplify the calculations involving (relative) homology and
cohomology below, and because we will eventually only be interested
in special Lagrangian {\it homology $3$-spheres} $N$, which have
$H_1(N,\R)=\{0\}$. The assumptions can be removed without great
difficulty.
\end{itemize}

Suppose Condition \ref{cs4cond} holds. Define $P=N_0\sm\Phi_0(B_\rho)$.
Then $P$ is a compact, nonsingular 3-manifold whose boundary is the
image under $\Phi_0$ of the intersection of $L_0$ with the sphere
${\cal S}^5_\rho$ in $\C^3$ of radius $\rho$. Now $L_0\cap{\cal S}^5_\rho$
is the orbit of $3^{-1/2}\rho(1,1,1)$ under the action of $G=\U(1)^2$
defined in \eq{cs3eq1}. So define $\iota:G\ra P$ by
\e
\iota:({\rm e}^{i\th_1},{\rm e}^{i\th_2})\mapsto
\ts\Phi_0\bigl(\frac{1}{\sqrt{3}}\rho{\rm e}^{i\th_1},
\frac{1}{\sqrt{3}}\rho{\rm e}^{i\th_2},\frac{1}{\sqrt{3}}
\rho{\rm e}^{-i\th_1-i\th_2}\bigr).
\label{cs4eq1}
\e
Then $\iota$ is a diffeomorphism $G\ra\pd P$.

\begin{lem} The inclusion $\iota:G\ra P$ induces a map 
$\iota_*:H_1(G,\Z)\ra H_1(P,\Z)$, which has~$\Ker(\iota_*)\cong\Z$. 
\label{cs4lem1}
\end{lem}

\begin{proof} As $H_1(G,\Z)\cong\Z^2$ and $\Ker(\iota_*)$ 
is a subgroup of $H_1(G,\Z)$, we see that $\Ker(\iota_*)$ is 
isomorphic to $0$, $\Z$ or $\Z^2$. But using Poincar\'e duality 
ideas for manifolds with boundary, we can show that the map 
$H_1(G,\R)\ra H_1(P,\R)$ must have image and kernel $\R$, and 
this forces~$\Ker(\iota_*)\cong\Z$.
\end{proof}

Using Poincar\'e duality for $P$ and $H_1(N_0,\R)=\{0\}$ one
can also show that $H_1(P,\R)\cong\R$ and $H_2(P,\R)=\{0\}$.
Write $Q_0=N_0\cap\Phi_0(\ovB_\rho)$. Then $Q_0$ is a singular
3-manifold with boundary $\pd Q_0=\pd P=\iota(G)$. Topologically
$Q_0$ is the cone on $\pd P$, and is contractible. We shall now
define homology classes $\ze,\chi$ and integers $k^1,k^2,k^3$
which will be important in what follows.

\begin{dfn} Let $\ze\in H_1(G,\Z)$ be a generator for
$\Ker(\iota_*)\cong\Z$. Then $\ze$ is unique up to sign. Now
$G=\U(1)^2$, so we may identify $H_1(G,\Z)\cong\Z^2$ in the
obvious way, such that the maps ${\cal S}^1\ra G$ given by
${\rm e}^{i\th}\mapsto({\rm e}^{i\th},1)$ and ${\rm e}^{i\th}\mapsto
(1,{\rm e}^{i\th})$ represent the classes in $H_1(G,\Z)$ identified
with $(1,0)$ and $(0,1)$ in $\Z^2$ respectively. Define $k^1,k^2,k^3\in\Z$
so that $\ze\in H_1(G,\Z)$ is identified with $(k^1,k^2)\in\Z^2$, and
$k^3=-k^1-k^2$. Then~$k^1+k^2+k^3=0$.

Let $\tau$ be a closed integral 1-chain in $\pd P$ with
$[\tau]=\iota_*(\ze)$ in $H_1(\pd P,\Z)$. Then $[\tau]=\iota_*(\ze)=0$
in $H_1(P,\Z)$, so there exists an integral 2-chain $\La$ in $P$ with
$\pd\La=\tau$. Also $\tau$ is a closed integral 1-chain in $\pd Q_0$
and $Q_0$ is contractible, so there exists an integral 2-chain $\Si$ in
$Q_0$ with $\pd\Si=\tau$. Thus $\Si-\La$ is an integral 2-chain without
boundary in $N_0\subset M$. Define $\chi=[\Si-\La]\in H_2(M,\R)$.
Then $\chi$ is a homology class in the image of $H_2(M,\Z)$ in
$H_2(M,\R)$. Since $H_2(P,\R)=H_2(Q_0,\R)=\{0\}$ this $\chi$
is independent of the choice of $\La,\Si$, and so is unique
up to the choice of sign of~$\ze$.
\label{cs4def1}
\end{dfn}

Note that as the chain $\Si-\La$ representing $\chi$ lies in $N_0$,
which is Lagrangian with respect to $\om_0$, we have $[\om_0]\cdot\chi=0$,
where $[\om_0]\in H^2(M,\R)$ is the {\it K\"ahler class} of $\om_0$. In
the rest of the section we shall consider separately the cases
$\chi\ne 0$ and~$\chi=0$.

\subsection{The case $\chi\ne 0$: desingularizing $N_0$}
\label{cs42}

Let $(M,J_0,\om_0,\Om_0)$, $N_0$ and $x$ be as above. We shall now
consider the question of when $N_0$ is the limit as $t\ra 0_+$ of a
family of compact, nonsingular SL 3-folds modelled on $L^a_t$ near
$x$. In general this cannot happen in the fixed almost Calabi--Yau
3-fold $(M,J_0,\om_0,\Om_0)$. Instead we need to suppose $\chi\ne 0$,
and extend $(M,J_0,\om_0,\Om_0)$ to a smooth family of almost
Calabi--Yau 3-folds $(M,J_t,\om_t,\Om_t)$ with $[\om_t]\cdot\chi=t$.
Here is our conjecture.

\begin{conj} Let Condition \ref{cs4cond} hold, and\/ $k^1,k^2,k^3$
and\/ $\chi$ be as in Definition \ref{cs4def1}. Suppose $\chi\ne 0$,
and\/ $(M,J_0,\om_0,\Om_0)$ extends to a smooth family
$\bigl\{(M,J_t,\om_t,\Om_t):t\in(-\ep,\ep)\bigr\}$ of almost
Calabi--Yau $3$-folds for some $\ep>0$, where~$[\om_t]\cdot\chi=t$.

Then there do not exist singular SL\/ $3$-folds $N_t$ in
$(M,J_t,\om_t,\Om_t)$ close to $N_0$ and of the same topological
type for $t\ne 0$. However, for some small\/ $\de\in(0,\ep]$,
whenever $a=1,2,3$, $\md{t}<\de$ and\/ $k^at>0$ there exists
a compact, nonsingular, embedded SL\/ $3$-fold $N^a_t$ in
$(M,J_t,\om_t,\Om_t)$ with phase ${\rm e}^{i\th_t}$, which
depends smoothly on $t$, and converges to $N_0$ as $t\ra 0$ in a
suitable sense, and near $x$ is locally modelled on~$L^a_{t/k^a\pi}$.

To be more precise, whenever $a=1,2,3$, $0<\md{t}<\de$ and\/ $k^at>0$,
there exists an embedding $\Phi^a_t:\ovB_\rho\ra M$ which depends
smoothly on $t$ and converges to $\Phi_0$ as $t\ra 0$, and satisfies
\begin{itemize}
\item[{\rm(a)}] $(\Phi^a_t)^*(N^a_t)=L^a_{t/k^a\pi}\cap\ovB_\rho$;
\item[{\rm(b)}] $(\Phi^a_t)^*(\om_t)=\om'$; and
\item[{\rm(c)}] $(\Phi^a_t)^*(J_t)\approx J'$ and\/ $(\Phi^a_t)^*(\Om_t)
\approx F_t{\rm e}^{i\th_t}\Om'$ for small\/ $t$, near $0\in\ovB_\rho$,
for some~$F_t>0$.
\end{itemize}
\label{cs4conj1}
\end{conj}

In the rest of this subsection we shall give a partial proof of
this conjecture, which justifies the assertion that $N_t$ does not
exist for $t\ne 0$ but that $N^a_t$ should exist when $k^at>0$, and
be locally modelled on $L^a_{t/k^a\pi}$, rather than $L^a_s$ for
some other $s>0$. To complete the proof will require essentially
the same analysis needed to solve Conjecture \ref{cs2conj2}, and
so once Conjecture \ref{cs2conj2} has been proved, completing the
conjecture above should be straightforward.

First, the nonexistence of singular SL 3-folds $N_t$ in
$(M,J_t,\om_t,\Om_t)$ for $t\ne 0$ close to $N_0$, and of the
same topological type. Suppose such an $N_t$ did exist. Then
$N_t$ is homeomorphic to $N_0$, and as in \S\ref{cs41} we can
construct a closed integral 2-chain $C$ in $N_t$ close to $\Si-\La$
in $N_0$. Then $[C]=\chi$ in the image of $H_2(M,\Z)$ in
$H_2(M,\R)$, as $\chi$ is a discrete object and is unchanged 
under small variations. But $\om_t\vert_C=0$, as $C$ lies in
$N_t$ which is Lagrangian with respect to $\om_t$. Hence
$[\om_t]\cdot\chi=0$, which contradicts the assumptions
$[\om_t]\cdot\chi=t$ and~$t\ne 0$.

Suppose that for some small $t\in(-\ep,\ep)$ there exists a compact
nonsingular SL 3-fold $N^a_t$ in $(M,J_t,\om_t,\Om_t)$, which is
close to $N_0$ in a suitable sense, and near $x$ is modelled on
$L^a_s$. That is, there should exist an embedding $\Phi^a_t:\ovB_\rho
\ra M$ satisfying $(\Phi^a_t)^*(N^a_t)=L^a_s\cap\ovB_\rho$ and parts
(b), (c) above.

Define $P^a_t=N^a_t\sm\Phi^a_t(B_\rho)$ and $Q^a_t=N^a_t\cup\Phi^a_t
(\ovB_\rho)$. Then $Q^a_t$ is the image under $\Phi^a_t$ of
$L^a_s\cap\ovB_\rho$. Provided $s<\rho^2$, which we may assume
as $s$ is small, $Q^a_t$ is diffeomorphic to ${\cal S}^1\t D^2$, and
$P^a_t$ is a compact, nonsingular 3-manifold with boundary $\pd P^a_t
=\pd Q^a_t\cong T^2$. Furthermore, $P^a_t$ is diffeomorphic to $P$ in
\S\ref{cs41}. (In our present situation this follows from the assumption
that $N^a_t$ is close to $N_0$, and in the situation of the conjecture
from the fact that $P^a_t$ depends smoothly on $t$ and converges to
$P$ as~$t\ra 0$.)

Now the boundary of $L^a_s\cap\ovB_\rho$ is a $G$-orbit, so by
picking a point in the boundary and using $\Phi^a_t$ we may define
a diffeomorphism $\iota^a_t:G\ra\pd P^a_t=\pd Q^a_t$, in the same
way that we defined $\iota$ in \S\ref{cs41}. Furthermore, the
diffeomorphism $P\cong P^a_t$ may be chosen to identify $\iota$
and~$\iota^a_t$.

Let $\ze$ be as in Definition \ref{cs4def1}, and let $\tau^a_t$ be a
closed integral 1-chain in $\pd P^a_t$ with $[\tau^a_t]=(\iota^a_t)_*
(\ze)$ in $H_1(\pd P^a_t,\Z)$. Then $[\tau^a_t]=(\iota^a_t)_*(\ze)=0$
in $H_1(P^a_t,\Z)$, so there exists an integral 2-chain $\La^a_t$ in
$P^a_t$ with $\pd\La^a_t=\tau^a_t$. In \S\ref{cs31} we defined a
holomorphic disc $D^a_s$ in $\C^3$ with boundary $\pd D^a_s=\ga^a_s$
in $L^a_s$. Thus $\Phi^a_s(\ga^a_s)$ is an oriented ${\cal S}^1$ in
$Q^a_t$. As $Q^a_t\cong{\cal S}^1\t D^2$ we have $H_1(Q^a_t,\Z)\cong\Z$,
and it is easy to see that $[\Phi^a_s(\ga^a_s)]$ generates~$H_1(Q^a_t,\Z)$.

Now $\tau^a_t$ is a closed 1-chain in $\pd Q^a_t$, and the integers
$k^1,k^2,k^3$ of Definition \ref{cs4def1} were chosen to ensure that
$[\tau^a_t]=(\iota^a_t)_*(\ze)=k^a[\Phi^a_s(\ga^a_s)]$ in $H_1(Q^a_s,\Z)$.
Therefore $\tau^a_t$ is homologous to $k^a\Phi^a_t(\ga^a_s)$ in
$Q^a_t$, so there exists an integral 2-chain $\Si^a_t$ in $Q^a_t$
with~$\pd\Si^a_t=\tau^a_t-k^a\Phi^a_t(\ga^a_s)$.

Consider the integral 2-chain $k^a\Phi^a_t(D^a_s)+\Si^a_t-\La^a_t$ in
$M$. As $\pd D^a_s=\ga^a_s$, $\pd\Si^a_t=\tau^a_t-k^a\Phi^a_t(\ga^a_s)$
and $\pd\La^a_t=\tau^a_t$ it is closed, and so defines a homology class
in the image of $H_2(M,\Z)$ in $H_2(M,\R)$. Now this class is exactly
$\chi$, as we constructed it by a small deformation of the construction
of $\chi$ in Definition \ref{cs4def1}, but $\chi$ is a discrete object
and so is unchanged by continuous deformations. Therefore we have
\begin{equation*}
[\om_t]\cdot\chi=k^a\int_{\Phi^a_t(D^a_s)}\om_t+\int_{\Si^a_t}\om_t-
\int_{\La^a_t}\om_t=k^a\int_{D^a_s}\om'=k^a\pi s,
\end{equation*}
using part (b) of the conjecture and the fact that $\Si^a_t$ and
$\La^a_t$ lie in $N^a_t$, which is Lagrangian with respect to~$\om_t$.

Now $[\om_t]\cdot\chi=t$ by definition, so we have shown that
$t=k^a\pi s$. But $s>0$, as $L^a_s$ is only defined for positive
$s$. Thus, if $t$ and $k^a$ are not both zero then $N^a_t$ can only
exist if $k^at>0$, and then $s=t/k^a\pi$, as we have to prove. The
case $t=k^a=0$ is excluded by the conjecture, and will be discussed
in \S\ref{cs43}. This concludes our partial proof of
Conjecture~\ref{cs4conj1}.

\subsection{The case $\chi\ne 0$: homology 3-spheres}
\label{cs43}

Next we prove that the 3-manifolds $N^a_t$ above are homology
3-spheres, and calculate the size of~$H_1(N^a_t,\Z)$.

\begin{prop} In the situation of Conjecture \ref{cs4conj1},
both\/ $H_1(N_0,\Z)$ and $H_1(N^a_t,\Z)$ are finite and
satisfy $\bmd{H_1(N^a_t,\Z)}=\md{k^a}\cdot\bmd{H_1(N_0,\Z)}$.
Thus, each $N^a_t$ is a (rational) homology $3$-sphere.
\label{cs4prop}
\end{prop}

\begin{proof} Let $a=1,2$ or 3 and $t$ satisfy $\md{t}<\de$ and
$k^at>0$, and let $N^a_t$ be as in Conjecture \ref{cs4conj1}. Suppose
also that $t$ is small enough that $\md{t}<\md{k^a}\pi\rho^2$, to
ensure that $Q^a_t\cong{\cal S}^1\t D^2$, and define $P^a_t$, $Q^a_t$
and $\iota^a_t$ as in \S\ref{cs42}. As the $N^a_t$ depend smoothly on
$t$ they are all diffeomorphic for fixed $a$, and so it is enough to
prove the proposition for sufficiently small~$t$.

Using the exact sequence
\begin{equation*}
H_1(\{x\},\Z)\ra H_1(N_0,\Z)\ra H_1(N_0;\{x\},\Z)\ra
H_0(\{x\},\Z){\buildrel\cong\over\longra}
H_0(N_0,\Z)
\end{equation*}
and $H_1(\{x\},\Z)=0$, we find that $H_1(N_0,\Z)\cong H_1(N_0;\{x\},\Z)$,
and by excision and the diffeomorphism $P\cong P^a_t$ we see that
$H_1(N_0;\{x\},\Z)\cong H_1(P^a_t;\pd P^a_t,\Z)\cong H_1(N^a_t;Q^a_t,\Z)$.
Thus $H_1(N^a_t;Q^a_t,\Z)\cong H_1(N_0,\Z)$. Recall that $H_1(N_0,\R)=
\{0\}$ by Condition \ref{cs4cond}. It is easy to show using this and
facts about manifold topology that $H_1(N_0,\Z)$ is finite.

As $Q^a_t$ is diffeomorphic to ${\cal S}^1\t\R^2$ we see that
$H_1(Q^a_t,\Z)$ is isomorphic to $\Z$, and is generated by
$[\Phi^a_t(\ga^a_{t/k^a\pi})]$. But the argument in \S\ref{cs42}
shows that $k^a\Phi^a_t(\ga^a_{t/k^a\pi})$ is homologous in $Q^a_t$
to $\tau^a_t$ in $\pd Q^a_t$, and $\tau^a_t$ is homologous to 0 in
$P^a_t$. Thus, the image of $\md{k^a}\,[\Phi^a_t(\ga^a_{t/k^a\pi})]$ in
$H_1(N^a_t,\Z)$ is zero. Furthermore, no smaller positive multiple of
$[\Phi^a_t(\ga^a_{t/k^a\pi})]$ can have zero image in $H_1(N^a_t,\Z)$,
because $\tau^a_t$ represents $\ze$, which by definition generates the
kernel of~$(\iota^a_t)_*$.

Now consider the exact sequence
\begin{equation*}
H_1(Q^a_t,\Z)\ra H_1(N^a_t,\Z)\ra H_1(N^a_t;Q^a_t,\Z)\ra
H_0(Q^a_t,\Z){\buildrel\cong\over\longra}
H_0(N^a_t,\Z).
\end{equation*}
We have shown above that $H_1(N^a_t;Q^a_t,\Z)\cong H_1(N_0,\Z)$, which
is finite, and that the image of $H_1(Q^a_t,\Z)$ in $H_1(N^a_t,\Z)$ is
isomorphic to the cyclic group $\Z_{\md{k^a}}$. Thus by exactness
$H_1(N^a_t,\Z)/\Z_{\md{k^a}}\cong H_1(N_0,\Z)$, and so $H_1(N^a_t,\Z)$
is finite with $\bmd{H_1(N^a_t,\Z)}=\md{k^a}\cdot\bmd{H_1(N_0,\Z)}$,
as we have to prove.

Since $H_1(N^a_t,\Z)$ is finite we have $b_1(N^a_t)=0$, so $b_2(N^a_t)=0$
by Poincar\'e duality. But $N^a_t$ is connected (this is part of our
definition of manifold) and oriented, as it is special Lagrangian. So
$N^a_t$ is by definition a rational homology 3-sphere.
\end{proof}

Our goal in this paper is to define an invariant of almost Calabi--Yau
3-folds by counting special Lagrangian homology 3-spheres in some
appropriate way. We can use the ideas above to draw some conclusions
about how to do this. We assume for the moment that Conjecture
\ref{cs4conj1} is true.

Because $k^1+k^2+k^3=0$ and the $k^a$ are not all zero as $\ze\ne 0$,
at least one $k^a$ is positive, and one negative. But the SL 3-fold
$N^a_t$ exists for small $t>0$ if and only if $k^a>0$, and for small
$t<0$ if and only if $k^a<0$. So, consider the following three cases:
\begin{itemize}
\item[(a)] Suppose $k_1,k_2>0$ and $k_3<0$. Then $N^1_t$ and $N^2_t$
exist as SL 3-folds for small $t>0$ but $N^3_t$ does not, and $N^3_t$
exists for small $t<0$ but $N^1_t,N^2_t$ do not.
\item[(b)] Suppose $k_1>0$, $k_2=0$ and $k_3<0$. Then $N^1_t$ exists
for small $t>0$, $N^3_t$ exists for small $t<0$, and $N^2_t$ does
not exist for any $t\ne 0$.
\item[(c)] Suppose $k_1>0$ and $k_2,k_3<0$. Then $N^1_t$ exists
for small $t>0$, and $N^2_t,N^3_t$ exist for small $t<0$.
\end{itemize}
These show that as we deform $(M,J_t,\om_t,\Om_t)$, changing the
K\"ahler class $[\om_t]$, it can happen that two SL homology
3-spheres disappear, and one reappears; or that one disappears,
and another reappears; or that one disappears, and two reappear.

In particular, this shows that the number of special Lagrangian
homology 3-spheres in $M$ is not invariant under deformations of
$(M,J,\om,\Om)$ changing the K\"ahler class $[\om]$, even if counted
with signs. So, simple counting of SL homology 3-spheres, even with
signs, is probably not the right thing to do. However, it is easy to
see using Proposition \ref{cs4prop} that the sum over SL homology
3-spheres $N$ of the weight $\bmd{H_1(N,\Z)}$ is {\it unchanged}\/
under transitions of the kind described in Conjecture~\ref{cs4conj1}.

For instance, in case (a) above, for small $t>0$ there exist two
SL homology 3-spheres $N^1_t,N^2_t$, with $\bmd{H_1(N^1_t,\Z)}=
k^1\bmd{H_1(N_0,\Z)}$ and $\bmd{H_1(N^2_t,\Z)}=k^2\bmd{H_1(N_0,\Z)}$,
and when $t<0$ there is one SL homology 3-spheres $N^3_t$ with
$\bmd{H_1(N^3_t,\Z)}=-k^3\bmd{H_1(N_0,\Z)}$. But as $-k^3=k^1+k^2$ we
see that $\bmd{H_1(N^1_t,\Z)}+\bmd{H_1(N^2_t,\Z)}=\bmd{H_1(N^3_t,\Z)}$,
so the sum of weights is the same for small $t>0$ and $t<0$. This
suggests that the appropriate thing to do is to count SL homology
3-spheres $N$ with weight $\bmd{H_1(N,\Z)}$, and we will argue this
in~\S\ref{cs7}.

We can now discuss the existence of SL 3-folds $N^a_t$ when $k^a=t=0$,
which was passed over in \S\ref{cs42}. The argument of \S\ref{cs42}
shows that for an SL 3-fold $N^a_t$ to exist in $(M,J_t,\om_t,\Om_t)$
that is modelled on $N_0$ away from $x$ and $L^a_s$ near $x$, we need
$t=k^a\pi s$. Thus, if $k^a=0$ then such SL 3-folds can only exist
when $t=0$. The argument above gives an exact sequence
\begin{equation*}
\Z=H_1(Q^a_t,\Z)\ra H_1(N^a_t,\Z)\ra H_1(N_0,\Z)\ra 0,
\end{equation*}
and shows that the generator of $H_1(Q^a_t,\Z)$ has order $\md{k^a}$
in $H_1(N^a_t,\Z)$. However, in this case $k^a=0$, so the map $\Z\ra
H_1(N^a_t,\Z)$ is {\it injective}. It follows that $H_1(N^a_t,\Z)$
is infinite, the product of $\Z$ with a finite group, and hence
that $b^1(N^a_t)=1$. So $N^a_t$ is not a homology 3-sphere, and
by Theorem \ref{cs2thm1} it is not isolated as an SL 3-fold, but
occurs in a moduli space of dimension~1.

Thus we are led to the following picture. If $k^a>0$ then there
exists a unique SL 3-fold $N^a_t$ in $(M,J_t,\om_t,\Om_t)$ for
small $t>0$. If $k^a<0$ then there exists a unique SL 3-fold
$N^a_t$ in $(M,J_t,\om_t,\Om_t)$ for small $-t>0$. And if $k^a=0$
then for each small $s>0$ there exists a unique SL 3-fold $N^a_0$
in $(M,J_0,\om_0,\Om_0)$ modelled near $x$ on $L^a_s$. So in each
case the SL 3-folds are indexed by a small positive parameter.

\subsection{The case $\chi=0$: stable SL singularities}
\label{cs44}

In \S\ref{cs32} we argued that SL singularities modelled on the SL
cone $L_0$ in $\C^3$ are of {\it index one}, and should occur in
codimension one in families of SL 3-folds in generic families of
almost Calabi--Yau 3-folds. However, such local calculations of
the `index' of singularities can sometimes give the wrong answer
in cases when global topological restrictions come into play.
This happens when $\chi=0$ in the situation of \S\ref{cs41},
and effectively the index of the singularities should be zero
rather than one in this case.

Let $\bigl\{(M,J_t,\om_t,\Om_t):t\in(-\ep,\ep)\bigr\}$ be a smooth
family of almost Calabi--Yau 3-folds deforming $(M,J_0,\om_0,\Om_0)$.
As $\chi=0$, we cannot assume that $[\om_t]\cdot\chi=t$, so the family
is essentially arbitrary. The argument in \S\ref{cs42} that there do
not exists singular SL 3-folds $N_t$ in $(M,J_t,\om_t,\Om_t)$ for
$t\ne 0$ close to $N_0$, and of the same topological type, is now
no longer valid, and the author conjectures that such $N_t$ do
exist for small~$t$.

What about the existence of nonsingular SL 3-folds resolving $N_0$?
The argument in \S\ref{cs42} shows that if for small $t$ there
exists a nonsingular SL 3-fold $N^a_{s,t}$ in $(M,J_t,\om_t,\Om_t)$
close to $N_0$ away from $x$ and modelled on $L^a_s$ near $x$, then
$k^a\pi s=[\om_t]\cdot\chi=0$. As $s>0$, this shows that such SL
3-folds $N^a_{s,t}$ cannot exist unless $k^a=0$. When $k^a=0$ such
$N^a_{s,t}$ can exist, and have $b^1(N^a_{s,t})=1$ as in \S\ref{cs43},
so they should occur in 1-parameter families in $(M,J_t,\om_t,\Om_t)$
by Theorem \ref{cs2thm1}, parametrized by~$s$.

We summarize our conclusions in the following conjecture, which should
be provable by the same methods as Conjecture~\ref{cs4conj1}.

\begin{conj} Let Condition \ref{cs4cond} hold, and\/ $k^1,k^2,k^3$
and\/ $\chi$ be as in Definition \ref{cs4def1}. Suppose $\chi=0$, and\/
$(M,J_0,\om_0,\Om_0)$ extends to a smooth family $\bigl\{(M,J_t,\om_t,
\Om_t):t\in(-\ep,\ep)\bigr\}$ of almost Calabi--Yau $3$-folds for some
$\ep>0$. Then for some $\de\in(0,\ep]$, $N_0$ extends to a smooth family
of compact, embedded, singular SL\/ $3$-folds $N_t$ in $(M,J_t,\om_t,\Om_t)$
for $t\in(-\de,\de)$, each of which has one singular point locally
modelled on~$L_0$.

If\/ $k^a=0$ then for small\/ $t\in(-\ep,\ep)$ and\/ $s>0$ there
exists a unique compact, nonsingular SL\/ $3$-fold\/ $N^a_{s,t}$ in
$(M,J_t,\om_t,\Om_t)$ close to $N_0$ away from $x$, and modelled on
$L^a_s$ near $x$. If\/ $k^a\ne 0$ there do not exist any such\/~$N^a_{s,t}$.
\label{cs4conj2}
\end{conj}

\section{Another model degeneration of SL 3-folds}
\label{cs5}

We now describe a family of explicit SL 3-folds $K_{\bs{\phi},A}$
in $\C^3$. This family was first found by Lawlor \cite{Lawl}, was made
more explicit by Harvey \cite[p.~139--140]{Harv}, and was discussed from
a different point of view by the author in \cite[\S 5.4(b)]{Joyc2}. Our
treatment is based on that of Harvey.

Let $a_1,a_2,a_3>0$, and define polynomials $p(x)$, $P(x)$ by
\begin{equation*}
p(x)=(1+a_1x^2)(1+a_2x^2)(1+a_3x^2)-1
\quad\text{and}\quad P(x)=\frac{p(x)}{x^2}.
\end{equation*}
Define real numbers $\phi_1,\phi_2,\phi_3$ and $A$ by
\begin{equation*}
\phi_k=a_k\int_{-\iy}^\iy\frac{\d x}{(1+a_kx^2)\sqrt{P(x)}}
\quad\text{and}\quad A=\frac{4\pi}{3}(a_1a_2a_3)^{-1/2}.
\end{equation*}
Clearly $\phi_k>0$ and $A>0$. But writing $\phi_1+\phi_2+\phi_3$
as one integral and rearranging gives
\begin{equation*}
\phi_1+\phi_2+\phi_3=\int_0^\iy\frac{p'(x)\d x}{(p(x)+1)\sqrt{p(x)}}
=2\int_0^\iy\frac{\d w}{w^2+1}=\pi,
\end{equation*}
making the substitution $w=\sqrt{p(x)}$. So $\phi_k\in(0,\pi)$
and $\phi_1+\phi_2+\phi_3=\pi$. It can be shown that this yields 
a 1-1 correspondence between triples $(a_1,a_2,a_3)$ with $a_k>0$, 
and quadruples $(\phi_1,\phi_2,\phi_3,A)$ with $\phi_k\in(0,\pi)$,
$\phi_1+\phi_2+\phi_3=\pi$ and~$A>0$.

For $k=1,2,3$ and $y\in\R$, define $z_k(y)$ by $z_k(y)=
{\rm e}^{i\psi_k(y)}\sqrt{a_k^{-1}+y^2}$, where 
\begin{equation*}
\psi_k(y)=a_k\int_{-\iy}^y\frac{\d x}{(1+a_kx^2)\sqrt{P(x)}}\,.
\end{equation*}
Now write $\bs{\phi}=(\phi_1,\phi_2,\phi_3)$, and define 
a submanifold $K_{\bs{\phi},A}$ in $\C^3$ by
\e
K_{\bs{\phi},A}=\bigl\{(z_1(y)x_1,z_2(y)x_2,z_3(y)x_3):
y\in\R,\; x_k\in\R,\; x_1^2+x_2^2+x_3^2=1\bigr\}.
\label{cs5eq1}
\e
Our next result comes from Harvey~\cite[Th.~7.78]{Harv}.

\begin{prop} The set\/ $K_{\bs{\phi},A}$ defined in \eq{cs5eq1} is an
embedded SL\/ $3$-fold in $\C^3$ diffeomorphic to ${\cal S}^2\t\R$. It
is asymptotically conical, with cone the union $\Pi_0\cup\Pi_{\bs\phi}$
of two special Lagrangian $3$-planes $\Pi_0,\Pi_{\bs\phi}$ given by
\e
\Pi_0=\bigl\{(x_1,x_2,x_3):x_j\in\R\bigr\},\;\>
\Pi_{\bs\phi}=\bigl\{({\rm e}^{i\phi_1}x_1,{\rm e}^{i\phi_2}x_2,
{\rm e}^{i\phi_3}x_3):x_j\in\R\bigr\}.
\label{cs5eq2}
\e
\label{cs5prop}
\end{prop}

Here is how to interpret the constant $A$. Using the above notation, define
\e
D_{\bs{\phi},A}=
\bigl\{(x_1{\rm e}^{i\phi_1/2},x_2{\rm e}^{i\phi_2/2},x_3{\rm e}^{i\phi_3/2}):
x_k\in\R,\; a_1x_1^2+a_2x_2^2+a_3x_3^2\le 1\bigr\}.
\label{cs5eq3}
\e
Then $D_{\bs{\phi},A}$ is a solid ellipsoid in $\C^3$, with boundary 
in $K_{\bs{\phi},A}$. The axes of $D_{\bs{\phi},A}$ have lengths 
$a_k^{-1/2}$ for $k=1,2,3$, and so the volume of $D_{\bs{\phi},A}$ 
is $A$. Furthermore, $D_{\bs{\phi},A}$ is calibrated with respect 
to $\Im(\Om_0)$. That is, we can regard $D_{\bs{\phi},A}$ as an
SL 3-fold of phase $i$, whereas $K_{\bs{\phi},A}$ has phase 1; so
that $D_{\bs{\phi},A}$ and $K_{\bs{\phi},A}$ are both special
Lagrangian, but of {\it perpendicular phase}.

We met a similar situation in \S\ref{cs3}. There we defined
an AC SL 3-fold $L^a_t$ depending on a real parameter $t>0$,
and a holomorphic 2-disc $D^a_t$ with boundary on $L^a_t$, and
area $\pi t$. Here we define an AC SL 3-fold $K_{\bs{\phi},A}$
depending on a real parameter $A>0$, and a special Lagrangian
3-disc $D_{\bs{\phi},A}$ of phase $i$, with boundary on
$K_{\bs{\phi},A}$ and area~$A$.

Next we apply the ideas of \S\ref{cs23} to $\Pi_0\cup\Pi_{\bs{\phi}}$
and the~$K_{\bs{\phi},A}$.

\begin{lem} In the notation of\/ \S\ref{cs23}, $\Pi_0\cup\Pi_{\bs{\phi}}$
and\/ $K_{\bs{\phi},A}$ satisfy $k=2$, $b^1(K_{\bs{\phi},A})=0$ and\/
$\lind(K_{\bs{\phi},A})=6$, and\/ $\Pi_0\cup\Pi_{\bs{\phi}}$ is rigid.
\label{cs5lem}
\end{lem}

\begin{proof} As $\Pi_0\cup\Pi_{\bs{\phi}}$ is a cone on two disjoint
copies of ${\cal S}^2$ we have $k=1$, and as $K_{\bs{\phi},A}$ is
diffeomorphic to ${\cal S}^2\t\R$ we have $b^1(K_{\bs{\phi},A})=0$.
There are two connected components $\Si_1,\Si_2$ of $\Si$, each of
which is isometric to the unit sphere ${\cal S}^2$ in $\R^3$ with
the round metric.

For both $\Si_1,\Si_2$ the only eigenvalue of $\De$ in $(0,6)$ is 2,
with multiplicity 3, and eigenvectors the restriction to ${\cal S}^2$
of linear functions on $\R^3$. So $\lind(K_{\bs{\phi},A})=3+3=6$.
Similarly, $\De$ has eigenvalue 6 with multiplicity 5 on each of
$\Si_1,\Si_2$, and eigenfunctions the restrictions to ${\cal S}^2$
of harmonic homogeneous quadratic polynomials on $\R^3$. As
$\dim\SU(3)=8$ and the subgroup $G_j$ of $\SU(3)$ preserving
$\Si_j$ is isomorphic to SO(3) with dimension 3, we see that
$\Pi_0\cup\Pi_{\bs{\phi}}$ is rigid as~$5=8-3$.
\end{proof}

Assuming Conjecture \ref{cs2conj2}, we see that in the sense
discussed in \S\ref{cs23}, singularities of SL 3-folds modelled
on the SL cones $\Pi_0\cup\Pi_{\bs{\phi}}$ should have {\it index
one}. That is, they should occur in codimension 1 in families of
SL 3-folds in a generic almost Calabi--Yau 3-fold, or more
generally in generic families of almost Calabi--Yau 3-folds.

Now $\Pi_0\cup\Pi_{\bs{\phi}}$ may be regarded as a singular SL cone
in $\C^3$ with isolated singular point at 0, but it is also the union
of two {\it nonsingular} special Lagrangian 3-planes. In the same way,
an embedded singular SL 3-fold $N$ in $(M,J,\om,\Om)$ with one singular
point $x$ modelled on $\Pi_0\cup\Pi_{\bs{\phi}}$ is either a nonsingular
{\it immersed}\/ SL 3-fold with one self-intersection point at $x$, or
the union of two nonsingular SL 3-folds $N^+,N^-$ which intersect at $x$,
and have {\it the same phase}. We shall study this second possibility.

\section{Topological behaviour of these singularities}
\label{cs6}

We shall study the following situation.

\begin{cond} Let $\ep>0$, and $\bigl\{(M,J_t,\om_t,\Om_t):t\in(-\ep,
\ep)\bigr\}$ be a smooth family of almost Calabi--Yau $3$-folds with
the same underlying 6-manifold $M$. Suppose that $N_0^+$ and $N_0^-$
are compact, embedded, nonsingular special Lagrangian homology 3-spheres
in $(M,J_0,\om_0,\Om_0)$ with the same phase ${\rm e}^{i\th}$, which
intersect transversely at one point $x$, such that the intersection
$N_0^+\cap N_0^-$ is positive in the sense of homology, using the
natural orientations on $N_0^\pm$ and~$M$.
\label{cs6cond}
\end{cond}

We can show using the results of \S\ref{cs22} that $N_0^\pm$
extend to families of SL 3-folds $N_t^\pm$ in~$(M,J_t,\om_t,\Om_t)$.

\begin{prop} Suppose Condition \ref{cs6cond} holds. Then there
exists $\de\in(0,\ep]$ such that\/ $N_0^+$ and\/ $N_0^-$ extend
to unique smooth families $\bigl\{N_t^\pm:t\in(-\de,\de)\bigr\}$ of
compact, nonsingular, embedded SL\/ $3$-folds in $(M,J_t,\om_t,\Om_t)$.
Each\/ $N_t^\pm$ is diffeomorphic to $N_0^\pm$ and homologous
to it in $H_3(M,\Z)$, and\/ $N_t^+$ and\/ $N_t^-$ intersect
transversely in one point, where the intersection $N_t^+\cap N_t^-$
is positive in the sense of homology.
\label{cs6prop1}
\end{prop}

\begin{proof} As $N_0^\pm$ are homology 3-spheres we have
$H^2(N_0^\pm,\R)=0$, so the condition $[\om_t\vert_{N_0^\pm}]=0$
in $H^2(N_0^\pm,\R)$ in Theorem \ref{cs2thm2} holds automatically.
Thus, it follows from Theorem \ref{cs2thm2} that for some
$\de\in(0,\ep]$ we can extend $N_0^\pm$ to smooth families
$N_t^\pm$ of compact, nonsingular, embedded SL 3-folds in
$(M,J_t,\om_t,\Om_t)$ for $t\in(-\de,\de)$. As these are
smooth, connected families the $N_t^\pm$ are clearly
diffeomorphic and homologous to~$N_0^\pm$.

Now $b^1(N_t^\pm)=b^1(N_0^\pm)=0$ as $N_0^\pm$ is a homology
3-sphere, so $N_t^\pm$ is rigid by Theorem \ref{cs2thm1}, that
is, it has no special Lagrangian deformations. Thus the families
$N_t^\pm$ are unique. To intersect transversely in one point is
an open condition on pairs of submanifolds, so by making $\de$
smaller if necessary we can suppose that $N_t^+$ and $N_t^-$
intersect transversely in one point, which will be a positive
intersection as~$[N_t^+]\cap[N_t^-]=[N_0^+]\cap[N_0^-]=1$.
\end{proof}

We shall use the following notation.

\begin{dfn} Let Condition \ref{cs6cond} hold, and use the notation of
Proposition \ref{cs6prop1}. Define $\chi^\pm=[N_0^\pm]$ in $H_3(M,\Z)$.
Then $[N_t^\pm]=\chi^\pm$ for $t\in(-\de,\de)$. Let the phase of
$N_t^\pm$ be ${\rm e}^{i\th_t^\pm}$. Then $\th_t^\pm$ is only defined
modulo $2\pi\Z$, but we fix it uniquely by requiring that $\th_0^\pm=\th$
and $\th_t^\pm$ should depend smoothly on $t$. Applying Lemma \ref{cs2lem}
to $N_t^\pm$ we see that
\e
[\Om_t]\cdot\chi^\pm=R_t^\pm{\rm e}^{i\th_t^\pm}
\quad\text{for $t\in(-\de,\de)$,}
\label{cs6eq1}
\e
for some $R_t^\pm>0$ depending smoothly on $t$. As $\th_0^+=\th_0^-=
\th$, by making $\de>0$ smaller if necessary we may assume that
\e
\bmd{\th_t^+-\th_t^-}<\pi\quad\text{for all $t\in(-\de,\de)$.}
\label{cs6eq2}
\e
Then $R_t^+{\rm e}^{i\th_t^+}+R_t^-{\rm e}^{i\th_t^-}\ne 0$ for
any $t\in(-\de,\de)$. Define $R_t>0$ and $\th_t$ by
\e
[\Om_t]\cdot(\chi^++\chi^-)=R_t^+{\rm e}^{i\th_t^+}+R_t^-
{\rm e}^{i\th_t^-}=R_t{\rm e}^{i\th_t}
\label{cs6eq3}
\e
for all $t\in(-\de,\de)$. Here $\th_t$ is only defined modulo
$2\pi\Z$, but we specify $\th_t$ uniquely by requiring that it
depend smoothly on $t$, and $\th_0=\th$. It is then easy to show that
$\th_t$ lies between $\th_t^+$ and $\th_t^-$ for all~$t\in(-\de,\de)$.
\label{cs6def}
\end{dfn}

One can show using a `simultaneous diagonalization' argument
that if $\Pi^+,\Pi^-$ are SL 3-planes with phase 1 in $\C^3$
intersecting only at 0 then there exists $B\in\SU(3)$ such
that $B\Pi^+=\Pi_0$ and $B\Pi^-=\Pi_{\bs\phi}$, where
\begin{equation*}
\Pi_0=\bigl\{(x_1,x_2,x_3):x_j\in\R\bigr\}\quad\text{and}\quad
\Pi_{\bs\phi}=\bigl\{({\rm e}^{i\phi_1}x_1,{\rm e}^{i\phi_2}x_2,
{\rm e}^{i\phi_3}x_3):x_j\in\R\bigr\}
\end{equation*}
for some $\phi_1,\phi_2,\phi_3\in(0,\pi)$, where $\phi_1+\phi_2+\phi_3=
\pi$ if $\Pi^+\cap\Pi^-$ is a positive intersection in the sense
of homology, and $\phi_1+\phi_2+\phi_3=2\pi$ if $\Pi^+\cap\Pi^-$ is a
negative intersection. Using this it is easy to prove:

\begin{prop} Suppose Condition \ref{cs6cond} holds. Then there
exists a complex linear isometry $\iota:\C^3\ra T_xM$ satisfying
\e
\iota^*(\Om_0)={\rm e}^{i\th}\d z_1\w\d z_2\w\d z_3,\quad
\iota^*(T_xN_0^+)=\Pi_0 \quad\text{and}\quad
\iota^*(T_xN_0^-)=\Pi_{\bs{\phi}},
\label{cs6eq4}
\e
where $\phi_1,\phi_2,\phi_3\!\in\!(0,\pi)$ with\/ $\phi_1\!+\!\phi_2\!+
\!\phi_3\!=\!\pi$, and\/ $\Pi_0,\Pi_{\bs\phi}$ are given in~\eq{cs5eq2}.
\label{cs6prop2}
\end{prop}

This shows that after adjusting the phase by ${\rm e}^{i\th}$,
the singular SL 3-fold $N_0^+\cup N_0^-$ is modelled on the SL
cone $\Pi_0\cup\Pi_{\bs{\phi}}$ in $\C^3$ considered in \S\ref{cs5}.
Thus, it is natural to consider whether there exist compact,
nonsingular SL 3-folds $N_t$ in $(M,J_t,\om_t,\Om_t)$ modelled
on $N_0^+\cup N_0^-$ away from $x$, and on $K_{\bs{\phi},A}$ near
$x$ for small $t\in(-\de,\de)$ and $A>0$. Here is our conjectural
answer to this question.

\begin{conj} In the situation of Condition \ref{cs6cond}, with the
notation defined above, for each small\/ $t\in(-\de,\de)$ with\/
$\th_t^+>\th_t^-$ there exists a unique compact, nonsingular
special Lagrangian $3$-fold\/ $N_t$ in $(M,J_t,\om_t,\Om_t)$
modelled on $N_0^+\cup N_0^-$ away from $x$, and on
$K_{\bs{\phi},A}$ near $x$, where $A>0$ is small, depends on
$t$, and is given approximately by
\e
A\approx R_t^+\sin(\th_t^+-\th_t)=R_t^-\sin(\th_t-\th_t^-).
\label{cs6eq5}
\e
Furthermore, $N_t$ is diffeomorphic to the connected sum
$N_0^+\# N_0^-$ and is also a homology $3$-sphere, with\/
$\bmd{H_1(N_t,\Z)}=\bmd{H_1(N_t^+,\Z)}\cdot\bmd{H_1(N_t^-,\Z)}$,
and\/ $[N_t]=\chi^++\chi^-\in H_3(M,\Z)$. If\/ $t$ is small
and\/ $\th_t^+\le\th_t^-$ then there do not exist any such
SL\/ $3$-folds~$N_t$.
\label{cs6conj}
\end{conj}

In the rest of the section we will give a partial proof of this
conjecture, justifying the assertion that $N_t$ should exist only
if $\th_t^+>\th_t^-$, the approximate value of $A$ in \eq{cs6eq5},
and the topological claims about $N_t$. We begin with the topology.
If the SL 3-fold $N_t$ exists then it is modelled on $N_0^+\cup
N_0^-$ away from $x$, and on $K_{\bs{\phi},A}$ near $x$. But
$K_{\bs{\phi},A}$ is a narrow `neck' diffeomorphic to~${\cal S}^2\t\R$.

Thus, topologically $N_t$ is made by removing the point $x$ from $N_0^+$
and $N_0^-$, and joining them together with a small ${\cal S}^2\t\R$
`neck'. That is, $N_t$ is the {\it connected sum} $N_0^+\# N_0^-$ of
$N_0^+$ and $N_0^-$. It follows that $N_t$ is a homology 3-sphere,
and $H_1(N_t,\Z)\cong H_1(N_t^+,\Z)\t H_1(N_t^-,\Z)$, so
that~$\md{H_1(N_t,\Z)}=\md{H_1(N_t^+,\Z)}\cdot\md{H_1(N_t^-,\Z)}$.

Therefore $b^1(N_t)=0$, and hence by Theorem \ref{cs2thm1} $N_t$ is
{\it rigid}, so that if it exists it should be unique. (This is not a
rigorous argument). It is also obvious that $N_t$ is homologous to
$N_0^+\cup N_0^-$, and so $[N_t]=[N_0^+]+[N_0^-]=\chi^++\chi^-$ in
$H_3(M,\Z)$. Equation \eq{cs6eq3} then gives $[\om_t]\cdot[N_t]=
R_t{\rm e}^{i\th_t}$. So by Lemma \ref{cs2lem} the phase of $N_t$
is~${\rm e}^{i\th_t}$.

Now if $N_t$ has phase ${\rm e}^{i\th_t}$ and is modelled on
$K_{\bs{\phi},A}$ near $x$, in a similar way to Conjecture
\ref{cs4conj1} we expect that for some small $\rho>0$ there should
exist local coordinates $\Phi_t:\ovB_\rho\ra M$ near $x$ satisfying
the conditions
\begin{itemize}
\item[{\rm(a)}] $\Phi_t^*(N_t)\approx K_{\bs{\phi},A}\cap\ovB_\rho$;
\item[{\rm(b)}] $\Phi_t^*(\om_t)\approx F_t\om'$ near $0\in\ovB_\rho$
for some $F_t>0$; and
\item[{\rm(c)}] $\Phi_t^*(J_t)=J'$ and\/ $\Phi_t^*(\Om_t)=
{\rm e}^{i\th_t}\Om'$.
\end{itemize}
Here we have chosen our coordinate system to be holomorphic and to
identify $\Om_t$ with ${\rm e}^{i\th_t}\Om'$, which means we can only
assume that $\phi_t^*(N_t)\approx K_{\bs{\phi},A}$ and~$\Phi_t^*(\om_t)
\approx F_t\om'$.

Equation \eq{cs5eq3} defined a special Lagrangian ellipsoid $D_{\bs{\phi},
A}$ in $\C^3$ with phase $i$ and boundary in $K_{\bs{\phi},A}$. As
$\Phi_t^*(N_t)\approx K_{\bs{\phi},A}\cap\ovB_\rho$ there exists a disc
$D_t$ in $\C^3$ close to $D_{\bs{\phi},A}$ with $\pd D_t\subset
\Phi_t^*(N_t)$. As $D_t$ is close to $D_{\bs{\phi},A}$ we have
\e
\int_{D_t}\Im\Om'\approx\int_{D_{\bs{\phi},A}}\Im\Om'=A,
\label{cs6eq6}
\e
as $D_{\bs{\phi},A}$ is calibrated with respect to $\Im\Om'$ and
has area~$A$.

Now $N_t\sm\Phi_t(\pd D_t)$ has two connected components, say $P_t^\pm$,
where $P_t^+$ is close to $N_0^+$ and $P_t^-$ to $N_0^-$. The closures
$\ovP_t^\pm$ are compact, oriented 3-manifolds with boundary. Careful
consideration of the orientations of $N_t$ and $D_t$ shows that as
3-chains in $M$ we have
\begin{equation*}
\pd(\ovP_t^+)=\Phi_t(\pd D_t)
\quad\text{and}\quad
\pd(\ovP_t^-)=-\Phi_t(\pd D_t).
\end{equation*}
Therefore $\ovP_t^\pm\pm\Phi_t(D_t)$ is an integral 3-chain in $M$
without boundary, and so defines a homology class in $H_3(M,\Z)$.
But clearly $\ovP_t^\pm\pm\Phi_t(D_t)$ is homologous to $N_0^\pm$,
and so~$[\ovP_t^\pm\pm\Phi_t(D_t)]=\chi^\pm$.

Consider the integral of the 3-form $\Im({\rm e}^{-i\th_t}\Om_t)$ over
$\ovP_t^\pm\pm\Phi_t(D_t)$. Since $N_t$ is special Lagrangian of phase
${\rm e}^{i\th_t}$, this form vanishes on $N_t$ and $P_t^\pm$. Thus
\begin{equation*}
\Im\bigl({\rm e}^{i\th_t}[\Om_t]\cdot\chi^\pm\bigr)=
\pm\int_{\Phi_t(D_t)}\Im({\rm e}^{-i\th_t}\Om_t)=
\pm\int_{D_t}\Im(\Om')\approx\pm A,
\end{equation*}
using part (c) above and equation \eq{cs6eq6}. Therefore
\begin{equation*}
A\approx \Im\bigl({\rm e}^{-i\th_t}[\Om_t]\cdot\chi^+\bigr)=
-\Im\bigl({\rm e}^{-i\th_t}[\Om_t]\cdot\chi^-\bigr).
\end{equation*}
Substituting in equation \eq{cs6eq1} yields \eq{cs6eq5}, as we have
to prove.

Now $\md{\th_t^+-\th_t^-}<\pi$ by \eq{cs6eq2}, and $\th_t$ lies in
between $\th_t^+$ and $\th_t^-$. Using these one can show that if
$\th_t^+>\th_t^-$ then $\th_t^+-\th_t$ and $\th_t-\th_t^-$ lie in
$(0,\pi)$ and the (approximate) value for $A$ in \eq{cs6eq5} is
positive, but if $\th_t^+\le\th_t^-$ then $\th_t^+-\th_t$ and
$\th_t-\th_t^-$ lie in $(-\pi,0]$ and the (approximate) value for $A$
is nonpositive. Thus, $\th_t^+>\th_t^-$ corresponds to the condition
that $A>0$, which is part of the definition of $K_{\bs{\phi},A}$, and
justifies our claim that $\th_t^+>\th_t^-$ should be the necessary and
sufficient condition for the existence of $N_t$ for small $t$. This
completes our partial proof of Conjecture~\ref{cs6conj}.

Note that Adrian Butscher \cite{Buts} has proved an analytic result
closely related to Conjecture \ref{cs6conj}, but for SL $m$-folds in
$\C^m$ satisfying certain boundary conditions rather than for compact
SL $m$-folds in almost Calabi--Yau $m$-folds. It seems likely that his
analysis can be extended to prove our conjecture.

\section{Counting SL homology 3-spheres}
\label{cs7}

The {\it Gromov--Witten invariants} of a symplectic manifold
$(M,\om)$ are defined by counting, with signs, the $J$-holomorphic
curves in $M$ satisfying certain homological conditions. A good
introduction to the subject is given by McDuff and Salamon
\cite{McSa}. One important feature of these invariants is that
they are very stable under deformations of the choice of almost
complex structure $J$ used to define them.

Now it seems a natural (but perhaps optimistic) question to ask
whether we can define similar invariants of almost Calabi--Yau
3-folds $(M,J,\om,\Om)$ by counting SL 3-folds $N$ in $M$ 
satisfying suitable homological conditions. Probably the simplest
such condition is to count 3-folds $N$ in some fixed homology
class in $H_3(N,\Z)$, and we will focus on this. We shall also
restrict our attention to (rational) homology 3-spheres, to get
zero-dimensional moduli spaces.

Thus we aim to define an invariant as follows. Let $(M,J,\om,\Om)$ be
a almost Calabi--Yau 3-fold, let $\de\in H_3(M,\Z)$, and let $S(\de)$
be the set of special Lagrangian homology 3-spheres $N$ in $M$ with 
$[N]=\de$. Suppose $S(\de)$ is finite, and define
\e
I(\de)=\sum_{N\in S(\de)}w(N),
\label{cs7eq1}
\e
where $w$ is a {\it weight function} taking values in a commutative
ring $R$, and $w(N)$ depends only on the topology of $N$. In this 
way we define a map $I:H_3(M,\Z)\ra R$, which we consider to be an
analogue of the Gromov--Witten invariants. For this invariant to be
interesting, we would like it to be stable under deformations of the
underlying almost Calabi--Yau 3-fold $(M,J,\om,\Om)$, or at least to
change in a predictable way as we make these deformations.

Therefore we need to know what can happen to special Lagrangian homology
3-spheres as we deform $(M,J,\om,\Om)$, and especially how they can
become singular, appear or disappear. Each such transition may change
the set of special Lagrangian homology 3-spheres, and thus the invariant 
$I(\de)$. For $I(\de)$ to be invariant or to transform nicely under 
these transitions, the weight function $w$ must satisfy some 
topological identities.

We have already described models for two such transitions in
\S\ref{cs4} and \S\ref{cs6}. We will calculate the conditions on 
$w$ for $I(\de)$ to be invariant under the change described in 
\S\ref{cs4}, and to transform in a certain simple way under the
change described in \S\ref{cs6}. It turns out that the weight 
function $w(N)=\bmd{H_1(N,\Z)}$ satisfies both of these conditions. 

We also propose corrections to \eq{cs7eq1} to include multiple covers
of special Lagrangian homology 3-spheres and special Lagrangian 3-folds
with stable singularities. We summarize our conclusions in Conjecture
\ref{cs7conj}, and then discuss the connections between our invariants
and String Theory.

\subsection{Invariance of $I(\de)$ under the transitions of \S\ref{cs4}}
\label{cs71}

In \S\ref{cs42}--\S\ref{cs43} we explained how, as we deform an
almost Calabi--Yau 3-fold $(M,J_t,\om_t,\Om_t)$ in a 1-parameter
family, three SL 3-folds $N^1_t,N^2_t,N^3_t$ can converge to the
same singular SL 3-fold $N_0$ with a $T^2$-cone singularity. This
happens on a hyperplane $[\om_t]\cdot\chi=0$ in the K\"ahler cone,
and $N^a_t$ exists as a nonsingular SL 3-fold in $(M,J,\om,\Om)$
if either $k^a>0$ and $[\om_t]\cdot\chi>0$, or $k^a<0$ and
$[\om_t]\cdot\chi<0$. It is easy to see that the condition for the
invariant $I(\de)$ given by \eq{cs7eq1} to be unchanged by this
transition is
\e
\sum_{a\in\{1,2,3\}:k^a>0}w(N^a_t)=
\sum_{a\in\{1,2,3\}:k^a<0}w(N^a_t).
\label{cs7eq2}
\e

Now if $N$ is a homology 3-sphere then $H_1(N,\Z)$ is a finite
group, so $\bmd{H_1(N,\Z)}$ is a positive integer. Take the 
commutative ring $R$ to be $\Z$, and define $w(N)=\bmd{H_1(N,\Z)}$.
Remarkably, it turns out that this weight function, perhaps the 
simplest nontrivial invariant of $N$ there is, satisfies~\eq{cs7eq2}.

\begin{prop} Define an integer-valued invariant\/ $w$ of
compact, nonsingular 3-manifolds $N$ by $w(N)=\bmd{H_1(N,\Z)}$
if\/ $H_1(N,\Z)$ is finite, and\/ $w(N)=0$ if\/ $H_1(N,\Z)$ is
infinite. Then \eq{cs7eq2} holds for all sets of\/ $3$-manifolds
$N^1_t,N^2_t,N^3_t$ constructed as in~\S\ref{cs42}.
\label{cs7prop1}
\end{prop}

The proof follows quickly from the material of~\S\ref{cs4}.

\subsection{Transformation of $I(\de)$ under the transitions of
\S\ref{cs6}}
\label{cs72}

We shall use the following notation.

\begin{dfn} Let $M$ be a compact 6-manifold, and suppose
$\chi^+,\chi^-\in H_3(M,\Z)$ are linearly independent over $\R$.
Define a subset $W(\chi^+,\chi^-)$ in $H^3(M,\C)$ by
\e
W(\chi^+,\chi^-)=\bigl\{\Phi\in H^3(M,\C):(\Phi\cdot\chi^+)
(\bar\Phi\cdot\chi^-)\in(0,\iy)\bigr\}.
\label{cs7eq3}
\e
Then $W(\chi^+,\chi^-)$ is a {\it real hypersurface} in $H^3(M,\C)$,
but not a hyperplane. Let $\Phi\in H^3(M,\C)$, and write $(\Phi\cdot
\chi^+)(\bar\Phi\cdot\chi^-)=R{\rm e}^{i\th}$, where $R\ge 0$ and
$\th\in(-\pi,\pi]$. Then $\Phi\in W(\chi^+,\chi^-)$ if $R>0$ and
$\th=0$. Let $\ep>0$ be small. We say that $\Phi$ lies on the {\it
positive side} of $W(\chi^+,\chi^-)$ if $R>0$ and $\th\in(0,\ep)$,
and on the {\it negative side} of $W(\chi^+,\chi^-)$ if $R>0$
and~$\th\in(-\ep,0)$.
\label{cs7def1}
\end{dfn}

Section \ref{cs6} studied a family of almost Calabi--Yau 3-folds
$(M,J_t,\om_t,\Om_t)$ containing SL homology 3-spheres $N^\pm_t$,
with $[N^\pm_t]=\chi^\pm$ in $H_3(M,\Z)$, such that $N_t^+\cup N_t^-$
is a transverse intersection at a single point, and positive in
homology. In the notation above, the phases ${\rm e}^{i\th^\pm_t}$
of $N^\pm_t$ are equal exactly when $[\Om_t]\in W(\chi^+,\chi^-)$,
and Conjecture \ref{cs6cond} says that as $[\Om_t]$ passes through
$W(\chi^+,\chi^-)$ from the negative to the positive side, a new SL
3-fold $N_t$ diffeomorphic to $N_t^+\# N_t^-$ is created,
with~$[N_t]=\chi^++\chi^-$.

Clearly, $N_t$ contributes $w(N_t)\ne 0$ to the invariant
$I(\chi^++\chi^-)$ defined in \eq{cs7eq1}. Therefore,
$I(\chi^++\chi^-)$ will take {\it different values} on the
positive and negative sides of $W(\chi^+,\chi^-)$. So, $I(\de)$
will {\it not\/} be unchanged under deformations of the almost
Calabi--Yau structure $(J,\om,\Om)$ on $M$ which change $[\Om]$
in $H^3(M,\C)$. Instead, we hope to arrange for $I(\de)$ to
transform according to some rules involving $[\Om]$ and the
values $I(\al)$ for other $\al$ in~$H_3(M,\Z)$.

To see what the simplest of these rules should be, let us
generalize the situation of \S\ref{cs6}, and suppose that
$[\Om_t]$ lies on the negative side of $W(\chi^+,\chi^-)$
for $t\in(-\de,0)$, on $W(\chi^+,\chi^-)$ for $t=0$, and on
the positive side of $W(\chi^+,\chi^-)$ for $t\in(0,\de)$,
but that $N_0^+$ and $N_0^-$ intersect transversely at
$k+l$ points $x_1,\ldots,x_k$ and $y_1,\ldots,y_l$, where
$N_0^+\cap N_0^-$ is positive at $x_j$ and negative at $y_j$,
in the sense of homology.

Then arguing as in \S\ref{cs6}, we find that for small
$t\in(0,\de)$ we expect $k$ distinct, immersed SL homology
3-spheres $N_t$ diffeomorphic to $N^+_0\# N^-_0$, the connected
sums of $N^+_0$ and $N^-_0$ at $x_1,\dots,x_k$, and for small
$t\in(-\de,0)$ we expect $l$ distinct, immersed SL homology
3-spheres $N_t$ diffeomorphic to $N^+_0\# N^-_0$, the connected
sums of $N^+_0$ and $N^-_0$ at $y_1,\dots,y_l$.

Hence, as $[\Om_t]$ passes through $W(\chi^+,\chi^-)$ going from
the negative to the positive side, we simultaneously create $k$ and
destroy $l$ immersed special Lagrangian copies of $N^+_0\# N^-_0$.
Suppose for simplicity that $N_0^\pm$ are the only SL homology 3-spheres
in their homology classes $\chi^\pm$. Then $I(\chi^+)=w(N_0^+)$ and
$I(\chi^-)=w(N_0^-)$. Write $I(\chi^++\chi^-)^+$ for the value of
$I(\chi^++\chi^-)$ at some point on the positive side of
$W(\chi^+,\chi^-)$, and $I(\chi^++\chi^-)^-$ for its value
at a nearby point on the negative side. Then we have
\e
I(\chi^++\chi^-)^+-I(\chi^++\chi^-)^-=(k-l)\cdot w(N_0^+\# N_0^-).
\label{cs7eq4}
\e

If the weight function $w$ satisfies the identity
\e
w(N_0^+\# N_0^-)=w(N_0^+)\cdot w(N_0^-)
\quad\text{for all homology 3-spheres $N_0^\pm$,}
\label{cs7eq5}
\e
where multiplication is in the commutative ring $R$, then 
\eq{cs7eq4} can be written
\e
I(\chi^++\chi^-)^+-I(\chi^++\chi^-)^-=(\chi^+\cap\chi^-)
\cdot I(\chi^+)\cdot I(\chi^-),
\label{cs7eq6}
\e
as $\chi^+\cap\chi^-=k-l$ and $w(N_0^+\# N_0^-)=w(N_0^+)\cdot w(N_0^-)=
I(\chi^+)\cdot I(\chi^-)$. Because of the bilinearity of the
r.h.s.\ of \eq{cs7eq6} in $I(\chi^+)$ and $I(\chi^-)$, it is 
easy to see that \eq{cs7eq6} should also hold when there are
finitely many SL homology 3-spheres in $\chi^\pm$, and not
just one in each.

Thus, \eq{cs7eq6} gives a formula for how we expect $I(\chi^++\chi^-)$
to change as we pass through the hypersurface $W(\chi^+,\chi^-)$. The
important thing about this formula is that the values of $I$ on the
positive side of $W(\chi^+,\chi^-)$ determine the values of $I$ on the
negative side, and vice versa. So although $I$ is not invariant under
deformations of $(M,J,\om,\Om)$ which alter the cohomology class $[\Om]$,
it transforms in a completely determined way, and that is more or less
as useful.

In \S\ref{cs71} we proposed the weight function 
$w(N)=\bmd{H_1(N,\Z)}$, for $N$ a homology 3-sphere. Now 
$H_1(N_0^+\# N_0^-,\Z)\cong H_1(N_0^+,\Z)\t H_1(N_0^-,\Z)$ as finite
groups, and so
\begin{equation*}
\bmd{H_1(N_0^+\# N_0^-,\Z)}=\bmd{H_1(N_0^+,\Z)}\cdot\bmd{H_1(N_0^-,\Z)} 
\end{equation*}
when $N_0^\pm$ are homology 3-spheres. Thus this weight function 
$w$ satisfies \eq{cs7eq5}, as we wish.

\subsection{Including multiple covers}
\label{cs73}

Let $N'$ be an embedded SL homology 3-sphere in an almost Calabi--Yau
3-fold $(M,J,\om,\Om)$ with $[N']=\chi'\in H_3(M,\Z)$, and $\iota:N\ra N'$
be a {\it finite cover} of $N'$ of degree $d>1$, and covering group $G$,
so that $\md{G}=d$. Suppose $N$ is also a homology 3-sphere. Then we can
regard $N$ as an {\it immersed\/} SL homology 3-sphere in $M$, with
$[N]=d[N']=d\chi'$ in $H_3(M,\Z)$. What contribution should $N$ make
to~$I(d\chi')$?

We claim that $N$ should contribute $w(N)/d$ to $I(d\chi')$, where $w$
is the weight function for embedded homology 3-spheres. Note that if $w$
takes values in $\Z$, then $w(N)/d$ takes values in $\Q$, so that our
invariant $I(d\chi')$ will be actually be a rational number. Here is
why. Let $N''$ be another SL homology 3-sphere in $M$ intersecting
$N'$ transversely at one point, with $N'\cap N''$ positive. Let $[N'']
=\chi''$ in $H_3(M,\Z)$. Deform $(M,J,\om,\Om)$ so that $[\Om]$ passes
through $W(\chi',\chi'')$ going from the negative to the positive side.
Since $[N]\cap[N'']=d$ in homology, we expect from \S\ref{cs72} to
create $d$ distinct new SL homology 3-spheres as we pass through 
$W(\chi',\chi'')$, all diffeomorphic to~$N\# N''$.

However, a little thought shows that we actually create only {\it one}
SL copy of $N\# N''$. Effectively this is because the internal symmetry
group $G$ of $N$ with $\md{G}=d$ identifies the $d$ copies of $N\# N''$,
so that they all give the same SL 3-fold. In order for \eq{cs7eq6} to
give the correct transformation law for $I(d\chi'+\chi'')$ across
$W(\chi',\chi'')$, we must give $N$ the weight $w(N)/d$ rather
than~$w(N)$.

\subsection{Including SL 3-folds with stable singularities}
\label{cs74}

In \S\ref{cs44} we described a class of SL 3-folds with {\it stable
singularities}. That is, in an almost Calabi--Yau 3-fold $(M,J,\om,\Om)$
one can have SL 3-folds $N$ with one or more $T^2$-cone singularities,
that persist under small deformations of $(J,\om,\Om)$. Should such
singular SL 3-folds also be counted in the invariant $I(\de)$, with
some appropriate topological weight?

I believe that the answer to this is yes. Furthermore, if the
weight for nonsingular SL 3-folds $N$ is $w(N)=\md{H_1(N,\Z)}$,
then the weight for singular SL 3-folds may in some circumstances
be {\it negative}. I have not yet sorted all the details of this
out, and hope to do so in a future paper. However, here are the
ideas which lead me to this conclusion.

In \S\ref{cs3}--\S\ref{cs4} and \S\ref{cs5}--\S\ref{cs6} we described
two kinds of `index one' singularity of SL 3-folds, and analysed their
topological effects. Now I know of at least two other kinds of `index
one' singularity, the first locally modelled on the SL 3-folds of
\cite[Ex.~7.4]{Joyc2}, and the second on the singularities described
in \cite[\S 6]{Joyc3}. The topological effects are complicated to
describe.

My calculations indicate that one thing that can happen with the first
kind of index one singularity, is that one can simultaneously create,
out of nothing, a nonsingular, immersed SL homology 3-sphere $N_0$ with
one point of self-intersection $x$, and an embedded, singular SL 3-fold
$N_1$ with one $T^2$-cone singularity $y$, of the kind considered in
\S\ref{cs44}, and with $k^a=0$ for some~$a=1,2,3$.

In fact, we also create a family of nonsingular, embedded, diffeomorphic
SL 3-folds $N_t$ for $t\in(0,1)$ with $b^1(N_t)=1$, which converge to
$N_0$ as $t\ra 0_+$ with local model $K_{\bs{\phi},A}$ near $x$, where
$A\ra 0_+$ as $t\ra 0_+$, and converge to $N_1$ as $t\ra 1_-$ with local
model $L^a_s$ near $y$, where $s\ra 0_+$ as $t\ra 1_-$. Then $N_t$ for
$t\in[0,1]$ all have the same homology class~$[N_t]=\chi\in H_3(M,\Z)$.

Now when it exists, $N_0$ contributes $w(N_0)$ to $I(\chi)$. In order
for the invariant $I$ to be unchanged under this kind of transition we
need $N_1$ to contribute $-w(N_0)$ to $I(\chi)$. It can be shown that
$H_1(N_0,\Z)\cong H_1(N_1,\Z)$. Thus, if the weight for $N_0$ is
$w(N_0)=\md{H_1(N_0,\Z)}$, as above, then the correct weight for
$N_1$ is~$w(N_1)=-\md{H_1(N_1,\Z)}$. 

So I conjecture that for SL 3-folds $N$ with one stable $T^2$-cone
singularity of the kind considered in \S\ref{cs44} and $k^a=0$ for
some $a=1,2,3$, the appropriate weight is $w(N)=-\md{H_1(N,\Z)}$.
I am not yet sure of the answer when the $k^a$ are all nonzero, or
there is more than one $T^2$-cone singularity.

\subsection{A preliminary conjecture}
\label{cs75}

I am now ready to formulate a first guess as to how to define 
an invariant counting special Lagrangian homology 3-spheres, and 
what its properties should be under deformations of the underlying
almost Calabi--Yau 3-fold.

\begin{conj} Let\/ $(M,J,\om,\Om)$ be a generic almost Calabi--Yau
$3$-fold. Then there exists $I:H_3(M,\Z)\ra\Q$ with the following
properties:
\begin{itemize}
\item[{\rm(a)}] For each\/ $\de\in H_3(M,\Z)$, let\/ $S(\de)$ be the
set of compact, immersed, possibly singular SL\/ $3$-folds $N$ in
$M$, with\/ $[N]=\de$ and\/ $b^1(N)=0$. Then $S(\de)$ is finite, and
\e
I(\de)=\sum_{N\in S(\de)}w(N),
\label{cs7eq7}
\e
where $w(N)\in\Q$ is a weight function depending only on the
topology of\/ $N$ and its immersion in~$M$.
\item[{\rm(b)}] Let\/ $N$ be a nonsingular immersed SL\/ $3$-fold
in $M$. Regard\/ $N$ as a nonsingular, compact\/ $3$-manifold with
immersion $\iota:N\ra M$. Define $G(N)$ to be the group of
diffeomorphisms $\phi:N\ra N$ with $\iota\circ\phi=\iota$. Then
$G(N)$ is finite, and the weight\/ $w(N)$ is
\e
w(N)=\frac{\bmd{H_1(N,\Z)}}{\bmd{G(N)}}\,.
\label{cs7eq8}
\e
\item[{\rm(c)}] $I$ is unchanged by continuous deformations of\/
$(M,J,\om,\Om)$ that change $[\om]$ but leave $[\Om]$ fixed, or
multiply $[\Om]$ by a nonzero complex number. Thus $I$ depends
only on the complex structure $J$ on $M$, and not on the metric~$g$.
\item[{\rm(d)}] When we deform $(M,J,\om,\Om)$ so that\/ $[\Om]$ passes
through one of the hypersurfaces $W(\chi^+,\chi^-)$ in $H^3(M,\C)$
given in Definition \ref{cs7def1}, the invariant\/ $I$ transforms 
according to a set of rules that we are not yet able to write down.

One of these rules should be that if\/ $\chi^+$ and\/ $\chi^-$ are
primitive elements of\/ $H_3(M,\Z)$ and\/ $I(\chi^++\chi^-)^\pm$ 
are the values of\/ $I(\chi^++\chi^-)$ at two nearby points on the 
positive and negative sides of\/ $W(\chi^+,\chi^-)$, then
\end{itemize}
\e
I(\chi^++\chi^-)^+-I(\chi^++\chi^-)^-=(\chi^+\cap\chi^-)
\cdot I(\chi^+)\cdot I(\chi^-).
\label{cs7eq9}
\e
\label{cs7conj}
\end{conj}

Here are some remarks on this conjecture. Firstly, in defining
Gromov--Witten invariants, $J$-holomorphic curves are counted
{\it with signs}. But I believe that there is no corresponding
way to define the sign of an SL 3-fold, and that SL 3-folds
should be counted without signs.

Secondly, one effect of assuming that $(M,J,\om,\Om)$ is {\it generic}
is that the only SL 3-folds that can exist in $M$ are those with
{\it stable singularities}. So there is no need to restrict to
SL 3-folds with stable singularities in defining~$S(\de)$.

Thirdly, to understand how $I(\de)$ transforms under
deformations, it is enough to consider smooth 1-parameter families
$\bigl\{(M,J_t,\om_t,\Om_t):t\in[0,1]\bigr\}$ of almost Calabi--Yau
3-folds, where $(M,J_0,\om_0,\Om_0)$ and $(M,J_1,\om_1,\Om_1)$ are
generic (so that the invariant $I$ is defined when $t=0$ or 1) and
the family itself is generic as a 1-parameter family.

Now by the ideas of \S\ref{cs23}, the only singularities of
SL 3-folds $N$ with $b^1(N)=0$ that we will encounter in such
families of almost Calabi--Yau 3-folds are singularities with
{\it index one}. Thus, provided the general picture of \S\ref{cs23}
is correct, to prove Conjecture \ref{cs7conj} we only need to know
about special Lagrangian singularities with index one, and higher
codimension singularities can be ignored.

Fourthly, in part (d) we have not given a complete set of rules for
the transformation of $I$ on the hypersurfaces $W(\chi^+,\chi^-)$.
Here is why. When we pass through the hypersurface $W(\chi^+,\chi^-)$
we expect to create or destroy new SL 3-folds with homology class
$\chi^++\chi^-$, which are connected sums of 3-folds with
homology classes $\chi^+$ and $\chi^-$. But this is only the 
simplest kind of transition which happens on~$W(\chi^+,\chi^-)$.

For instance, if $N_1^+,N_2^+\in S(\chi^+)$ and $N^-\in S(\chi^-)$,
then on $W(\chi^+,\chi^-)$ we may create a new SL homology 3-sphere
with homology class $2\chi^++\chi^-$, diffeomorphic to the triple
connected sum $N_1^+\# N^-\# N_2^+$. So there should be some change
to $I(2\chi^++\chi^-)$ on $W(\chi^+,\chi^-)$. Similarly, if $a,b$
are positive integers then we can try and take a multiple connected
sum of $a$ elements of $S(\chi^+)$ and $b$ elements of $S(\chi^-)$ to
get a new SL homology 3-sphere with homology class~$a\chi^++b\chi^-$.

However, there is a problem: can we include an SL 3-fold in such a
connected sum more than once, and if so, how? The answer to this
appears to be rather complex, and I do not yet understand it, which
is why I'm not ready to write down a full set of transformation
rules for~$I$.

\subsection{Relationships with String Theory}
\label{cs76}

I now want to argue that the invariant $I$ postulated above should
have an interpretation in String Theory, and may fit into a piece
of the Mirror Symmetry story for Calabi--Yau 3-folds which is not
understood at present. There are two main reasons for this:
\begin{itemize}
\item[(a)] The invariant $I$ counts objects which are significant
in String Theory, namely {\it isolated\/ $3$-branes}; and
\item[(b)] Mirror Symmetry strongly suggests that for a Calabi--Yau
3-fold $X$ there should be discrete, $\Q$-valued invariants defined 
on $H_3(X,\Z)$ or something like it, which are related under the
mirror transform to the Gromov--Witten invariants of the mirror
manifold $Y$. But no such invariants are known at present.
\end{itemize}

We first discuss reason (a). In String Theory, SL 3-folds correspond
roughly to physical objects called 3-{\it branes}. But a 3-brane is
not just a 3-dimensional submanifold $N$; it also carries with it a
complex line bundle over $N$ with a flat $\U(1)$-connection. (For a
discussion of this, see for instance Strominger, Yau and Zaslow
\cite{SYZ}.) We will call a 3-brane {\it isolated} if it admits no
deformations, which happens when $N$ is a rational homology 3-sphere.

If $N$ is a compact 3-manifold, then flat $\U(1)$-connections on 
$N$ are equivalent to group homomorphisms $H_1(N,\Z)\ra\U(1)$. But 
when $H_1(N,\Z)$ is finite, it is easy to show using the theory of 
finite abelian groups that the number of group homomorphisms
$H_1(N,\Z)\ra\U(1)$ is exactly $\md{H_1(N,\Z)}$. Hence, if $N$
is a special Lagrangian homology 3-sphere, then there are exactly
$\md{H_1(N,\Z)}$ flat $\U(1)$-connections over $N$, and so $N$
gives rise to $\md{H_1(N,\Z)}$ isolated 3-branes.

Therefore, for the case of nonsingular, embedded SL 3-folds,
{\it the invariant\/ $I(\de)$ discussed above counts the number
of isolated\/ $3$-branes in the homology class} $\de$. So it is
a natural thing to count from the String Theory point of view,
although the author formulated Conjecture \ref{cs7conj} without
knowing this.

Next we discuss reason (b). The following basic outline of
Mirror Symmetry in String Theory is by now well known, and
is described for instance in Greene and Plesser \cite{GrPl}.
To a pair $(X,S)$, where $X$ is a Calabi--Yau 3-fold and $S$
some `extra structure' which we will not worry about, a physicist
associates a Super Conformal Field Theory (SCFT). There are
believed to exist `mirror pairs' $(X,S)$ and $(Y,T)$
whose SCFT's are isomorphic under a simple involution of the
SCFT structure.

If $(X,S)$ and $(Y,T)$ are such a mirror pair,
then $H^{2,1}(X)\cong H^{1,1}(Y)$, and certain cubic
forms $I_X^{2,1}$ on $H^{2,1}(X)$ and $I_{Y}^{1,1}$
on $H^{1,1}(Y)$ agree. Here $I_X^{2,1}$ has a simple
definition, depending on the variation of complex structures
on $X$. However, $I_{Y}^{1,1}$ has a complicated
definition, involving an infinite sum over homology classes
$\al$ in $H_2(Y,\Z)$ of the `number of rational curves'
in $Y$ with homology class $\al$, multiplied by complex
functions of a standard form.

Now, here is the important point. If we accept the mirror
conjecture, then $I_X^{2,1}$ can be written as essentially the
sum of a power series with integer (or rational) coefficients.
{\it Is there a way of understanding, purely in terms of\/ $X$
and without introducing the mirror $Y$, why these numbers
should be integers, and what they mean?}

The author hopes that this question can be answered as follows.
There should be a way to write $I_X^{2,1}$ as a sum over finite
collections of elements of $H_3(X,\Z)$ of integer or rational
invariants similar to the invariant $I$ described in \S\ref{cs75},
multiplied by complex (holomorphic, in a suitable sense) functions
of $[\Om]$ of a standard form. As $[\Om]$ passes through a
hypersurface $W(\chi^+,\chi^-)$, these complex functions may change
discontinuously to compensate for changes in $I$, so that $I_X^{2,1}$
remains continuous. The reason the numbers in the series are integers
(or rationals) is then that they count collections of SL 3-folds
satisfying certain conditions.

\end{document}